\documentclass[aps,a4paper,twocolumn]{revtex4}

\usepackage{amsmath}
\usepackage{graphicx}
\usepackage{subfigure}
\usepackage[usenames,dvipsnames]{color}
\definecolor{darkblue}{RGB}{0,0,196}
\usepackage[colorlinks=true,linkcolor=darkblue,citecolor=darkblue,urlcolor=darkblue]{hyperref}
\usepackage{setspace}
\usepackage{multirow}
\usepackage{hyperref}
\usepackage{xcolor}
\hypersetup{
  colorlinks   = true, 
  urlcolor     = blue, 
  linkcolor    = blue, 
  citecolor   = blue 
}
\usepackage{graphicx}
\usepackage{amsmath,bbm}
\usepackage{amssymb,bm}
\usepackage{yfonts}
\usepackage{comment}
\usepackage[normalem]{ulem}
\begin{document}

\title{Probing strangeness with event topology classifiers in pp collisions at energies available at the CERN Large Hadron Collider with the rope hadronization mechanism in PYTHIA}
\author{Suraj Prasad$^1$}
\author{Bhagyarathi Sahoo$^1$}
\author{Sushanta Tripathy$^2$}
\author{Neelkamal Mallick$^1$}
\author{Raghunath Sahoo$^1$\footnote{Corresponding Author: Raghunath.Sahoo@cern.ch}}

\affiliation{$^1$Department of Physics, Indian Institute of Technology Indore, Simrol, Indore 453552, India}
\affiliation{$^2$CERN, CH 1211, Geneva 23, Switzerland}

\date{\today}

\begin{abstract}

In relativistic heavy-ion collisions, the formation of a deconfined and thermalized state of partons, known as quark-gluon plasma (QGP), leads to enhanced production of strange hadrons in contrast to proton-proton (pp) collisions, which are taken as baseline. This observation is known as strangeness enhancement in heavy-ion collisions and is considered one of the important signatures that can signify the formation of QGP. However, in addition to strangeness enhancement, recent measurements hint at observing several heavy-ion-like features in high multiplicity pp collisions at energies available at the CERN Large Hadron Collider. Alternatively, event shape observables, such as transverse spherocity, transverse sphericity, charged particle flattenicity, and relative transverse activity classifiers, can fundamentally separate hard interaction-dominated jetty events from soft isotropic events. These features of event shape observables can probe the observed heavy-ion-like features in pp collisions with significantly reduced selection bias and can bring all collision systems on equal footing. In this article, we present an extensive summary of the strange particle ratios to pions as a function of different event classifiers using the PYTHIA~8 model with color reconnection and rope hadronization mechanisms to understand the microscopic origin of strangeness enhancement in pp collisions and also prescribe the applicability of these event classifiers in the context of strangeness enhancement. Charged particle flattenicity is found to be most suited for the study of strangeness enhancement, and it shows a quantitative enhancement similar to that seen for the analysis based on the number of multi-parton interactions.

\end{abstract}
  
\maketitle


\section{Introduction}
\label{intro}

In heavy-ion collisions at the CERN Large Hadron Collider (LHC) and BNL Relativistic Heavy-Ion Collider (RHIC), sufficiently high energy density and temperature are reached where nuclear matter goes through a transition to the deconfined state of quarks and gluons, known as quark-gluon plasma (QGP). Enhanced production of particles with strangeness in heavy-ion collisions with respect to pp collisions, known as strangeness enhancement, was originally proposed as an indirect signature of the formation of QGP~\cite{Rafelski:1982pu, Koch:1986ud, Rafelski:1991rh}. Also, the study of strangeness production gives an insight into quantum chromodynamics (QCD). Before the collisions, the colliding nuclear matter contained only up and down quarks, and strange quarks were not present as valence quarks. In the early state after the collision, the strange quarks are produced via hard perturbative processes via flavor creation ($q\bar{q} \rightarrow s\bar{s}$), gluon fusion ($gg \rightarrow s\bar{s}$), and flavor excitation ($gs \rightarrow gs$, $qs \rightarrow qs$). These processes are responsible for the strange particle production with higher transverse momenta after the subsequent hadronization. The non-perturbative processes dominate in the low transverse momenta for the production of strange hadrons. 

In heavy-ion collisions, the abundances of strange particles relative to pions (lightest meson with up and down quarks) remain nearly constant with a change of collision centrality and collision energy~\cite{ALICE:2013xmt, ALICE:2016fzo}. In elementary collisions like $e^{+}e^{-}$ and pp collisions, based on string fragmentation models, the production of strange hadrons is significantly suppressed relative to hadrons, with only up and down quarks as strange quarks are heavier and less likely to be produced thermally ($T<m_{\rm s}$), in these hadronic and elementary collisions. 
Recent measurements of high multiplicity pp collisions at LHC energies have revealed that such systems exhibit features similar to heavy-ion collisions, such as long-range near-side angular correlation~\cite{CMS:2010ifv}, non-zero $v_2$ coefficients~\cite{CMS:2016fnw}, mass-ordering in hadron $p_{T}$-spectra and characteristic modifications of baryon-to-meson ratios~\cite{CMS:2012xvn}, and enhanced production of multi-strange hadrons~\cite{ALICE:2016fzo}.  The enhanced production of strangeness with respect to pion yields is in contrast to the traditional belief of such production being achievable in ultra-relativistic nucleus-nucleus collisions~\cite{ALICE:2016fzo}. These measurements point towards a common underlying physics mechanism across collision systems and energies~\cite{WA97:1999uwz, NA57:2010tnk, NA57:2006aux, NA49:2004irs, NA49:2008ysv, STAR:2003jis, STAR:2008inc, STAR:2006egk, STAR:2007cqw, ALICE:2013xmt, ALICE:2013wgn, ALICE:2015mpp, ALICE:2018pal, ALICE:2020nkc}. This enhancement feature is described in the canonical thermal phase space suppression picture~\cite{Hamieh:2000tk, Redlich:2001kb}. However, other indirect signatures of QGP, such as jet quenching, have been reported to be absent in small collision systems~\cite{Nagle:2018nvi}. Such behavior is quite challenging for currently popular event generators with relevant and contrasting physics models, like PYTHIA~8 and EPOS-LHC~\cite{Pierog:2013ria}, into reproducing all the observed behavior simultaneously for small collision systems. Thus, the QGP-like behavior in small systems is still unclear; its microscopic origin is yet to be understood. Several theoretical models have been proposed to explain such behavior in small systems. One such explanation comes from the multi-parton interactions (MPI)-based picture with color reconnection (CR) and ropes inherited in the PYTHIA 8 model. This model can describe some of the QGP-like behavior by including phenomenological final state hadronization mechanisms such as color reconnection~\cite{Sjostrand:2014zea, OrtizVelasquez:2013ofg}, rope hadronization (RH)~\cite{Bierlich:2014xba} and string shoving~\cite{Bierlich:2016vgw}. However, MPI can not be accessible directly in experiments. Thus, several measurements have been performed in experiments as a function of final state charged particle multiplicity, which shows a significant correlation with MPI based on Monmte Carlo (MC) studies. However, measurements by event selections based only on mid-rapidity multiplicity have shown a stronger than linear increase of high-$p_{\rm T}$ particles in high multiplicity (HM) collisions relative to the yield in minimum-bias (MB) pp collisions~\cite{ALICE:2016fzo}. This indicates a selection bias towards local fluctuations of choosing only hard pp collisions. To reduce such biases, the event selection is performed in different pseudorapidity intervals with respect to the observable of interest. However, it is found that such measurements are still biased by the hard processes at high-$p_{\rm T}$~\cite{ALICE:2022qxg}. Thus, such selection biases in measurements hinder the search for the origin of QGP-like behavior in small collision systems. 
To pinpoint the origin of these phenomena with significantly reduced selection bias and to bring all collision systems in equal footings, along with charged particle multiplicity ($N_{\rm{ch}}$), lately several event classifiers such as transverse spherocity ($S_{0}$) ~\cite{Cuautle:2014yda, Cuautle:2015kra, Ortiz:2017jho, Banfi:2010xy, Khuntia:2018qox, ALICE:2023bga} and transverse sphericity ($S_{\rm T}$) ~\cite{Hanson:1975fe,ALICE:2012cor, CMS:2018mdd} have been adopted from the usage in $e^{+}e^{-}$ collisions since late 1970s. Also, new types of event shape observables have been constructed such as relative transverse activity classifiers $R_{\rm T}$, $R_{\rm T}^{\rm min}$, and $R_{\rm T}^{\rm max}$~\cite{Martin:2016igp,Bencedi:2021tst,Palni:2020shu} and charged particle flattenicity ($\rho_{\rm ch}$)~\cite{Ortiz:2022zqr,ALICE:2024vaf,Ortiz:2022mfv} to reduce the sensitivity to hard processes compared to the classifiers observed based only on charged particle multiplicity. The explorations using the above-mentioned observables have been performed extensively in experiments as well as on the phenomenological front. 

Various phenomenological models based on the string fragmentation process of hadronization in QCD form color ropes along with the CR mechanism qualitatively describe the strangeness enhancement at the LHC energies~\cite{Bierlich:2014xba, Bierlich:2018lbp, Flensburg:2011kk}. The QCD-inspired model based on Muellers's dipole formulation, called DIPSY rope~\cite{Bierlich:2015rha} qualitatively describes the strangeness enhancement feature for $\Lambda$ and $K_S^{0}$, while this model underestimates the production of $\Omega$ and $\Xi$ hyperons~\cite{ALICE:2016fzo}. Furthermore, a realistic MC model, called HERWIG~7~\cite{Duncan:2018gfk} based on the cluster mechanism of hadronization incorporates a new mechanism of color reconnection that describes the strangeness enhancement data quite well. In this model, the heavier hadrons, mostly baryons, are produced geometrically, and it incorporates the non-perturbative gluon splitting to produce more $s\bar{s}$ to explain the strange production.  On the other hand, another cluster reconnection model was introduced, which allows the reconnections between the baryonic and mesonic clusters that led to a better agreement with the data for the production of strange baryon~\cite{Gieseke:2017clv}. Recently, the effect of the range of color reconnection on strangeness enhancement is investigated in PYTHIA~8 considering an MPI-based CR picture. Further, a comparison of MPI-based CR and QCD-based CR is explored including color ropes~\cite{Hushnud:2023mgy}. Moreover, PYTHIA~8 with specific tunes of RH and CR mechanism predicts the strange hadrons to pion ratios quite well as a function of charged particle multiplicity in pp collisions at  $\sqrt{s}$ = 7 and 13  TeV ~\cite{ALICE:2018pal}. Thus, to understand the microscopic origin of strangeness enhancement in pp collisions in more detail, the study of the strangeness enhancement as a function of other existing event shape classifiers is necessary to get an insight into the applicability of event classifiers. Therefore, in this paper, we explore the strangeness enhancement as a function of various event shape observables such as MPI, transverse sphericity, transverse spherocity, relative transverse activity classifiers $R_{\rm T}$, and charged particle flattenicity $\rho_{\rm ch}$ using CR and RH mechanism of PYTHIA~8. Furthermore, to have a better understanding of the color ropes mechanism in the context of strangeness enhancement as a function of various event shape classifiers, we compare the baseline measurements of PYTHIA~8 without considering the effects of color ropes and color reconnection mechanisms in this study. \\ 

In PYTHIA~8, the hadron production occurs via the incoherent break-up of the strings, which exhibit a constant energy density. This keeps the strange particle ratios the same at different multiplicity events. However, in the rope hadronization framework, color ropes are incorporated in the PYTHIA~8 model, which involves different physical processes during the hadronization process. In the RH framework, the description of the interactions among the overlapping strings is given in two steps. Firstly, it allows the nearby strings to shove each other with an interaction potential. Then, during the hadronization process, the color charges at string endpoints and in gluon kinks can act together coherently to form a rope. This leads to an increase in the effective string tension in the events having large partonic interactions, which explains the strangeness enhancement feature in PYTHIA~8. Since, event classifiers such as $S_0$, $S_0^{p_{\rm T}=1}$, $S_{\rm T}$, $R_{\rm T}$ and $\rho_{\rm ch}$ are correlated with the number of multi-partonic interactions, this gives us an opportunity to perform the extensive study of event shape dependence of strangeness production using the various tunes of PYTHIA~8. In addition to their correlation with MPI, in this study, we discuss their coverages in terms of charged particle multiplicity and different biases present in each of these event shape observables in different physics tunes of PYTHIA~8. The studies presented here can collectively shed light on their applicability to study strangeness production in pp collisions at the LHC energies.

The paper is organized as follows. We start with a brief introduction in Sec.~\ref{intro}. The discussion on the event generation methodology and definitions of event classifiers are given in Sec.~\ref{pythiasphero}. The results are discussed in Sec.~\ref{results}, and Sec.~\ref{summary} provides a summary and outlook of the study.

\section{Event generation and methodology}
\label{pythiasphero}

In this section, we discuss the event generation using the Monte Carlo event generator PYTHIA~8, specific tunes used in this study, and analysis methodology. We start the description with the pQCD-inspired PYTHIA~8 model and then discuss different event classifiers such as charged particle multiplicity ($N_{\text{ch}}$), transverse sphericity ($S_{T}$), transverse spherocity  ($S_{0}$), and relative transverse activity classifier ($R_{T}$), and charged particle flattencity ($\rho_{\text{ch}}$).

\subsection{PYTHIA~8}
\label{sec:pythia}
PYTHIA is a widely used perturbative QCD-inspired Monte Carlo event generator to simulate hadronic, leptonic, and heavy-ion collisions with emphasis on physics related to small collision systems like pp collisions \cite{Sjostrand:2007gs}. PYTHIA~8 involves soft and hard QCD processes and contains the libraries for initial and final state parton showers, multiple parton-parton interactions, beam remnants, string fragmentation, and particle decays. A detailed explanation of all physics processes involved in PYTHIA~8 can be found in Ref.~\cite{manual}.

\begin{table}
\begin{tabular}{c c c c c c c c c c c c}
\hline \hline
\multicolumn{10}{c}{Rope Hadronization}& Values\\
\hline \hline
\multicolumn{10}{c}{ Ropewalk:RopeHadronization} &on\\
\multicolumn{10}{c}{ Ropewalk:doShoving} &on \\
\multicolumn{10}{c}{ Ropewalk:doFlavour} &on\\
\multicolumn{10}{c}{ Ropewalk:r0} &0.5\\
\multicolumn{10}{c}{ Ropewalk:m0} &0.2\\
\multicolumn{10}{c}{ Ropewalk:beta} &1.0 \\
\multicolumn{10}{c}{ Ropewalk:tInit} &1.0\\
\multicolumn{10}{c}{ Ropewalk:deltat} &0.05\\
\multicolumn{10}{c}{ Ropewalk:tShove} &10.0\\
\hline \hline
\end{tabular}
   \caption[p]{The parameter values of the PYTHIA rope hadronization model used with color reconnection mechanism.}
\label{table:rope}
 \end{table}

 \begin{figure}
    \centering
    \includegraphics[scale=0.4]{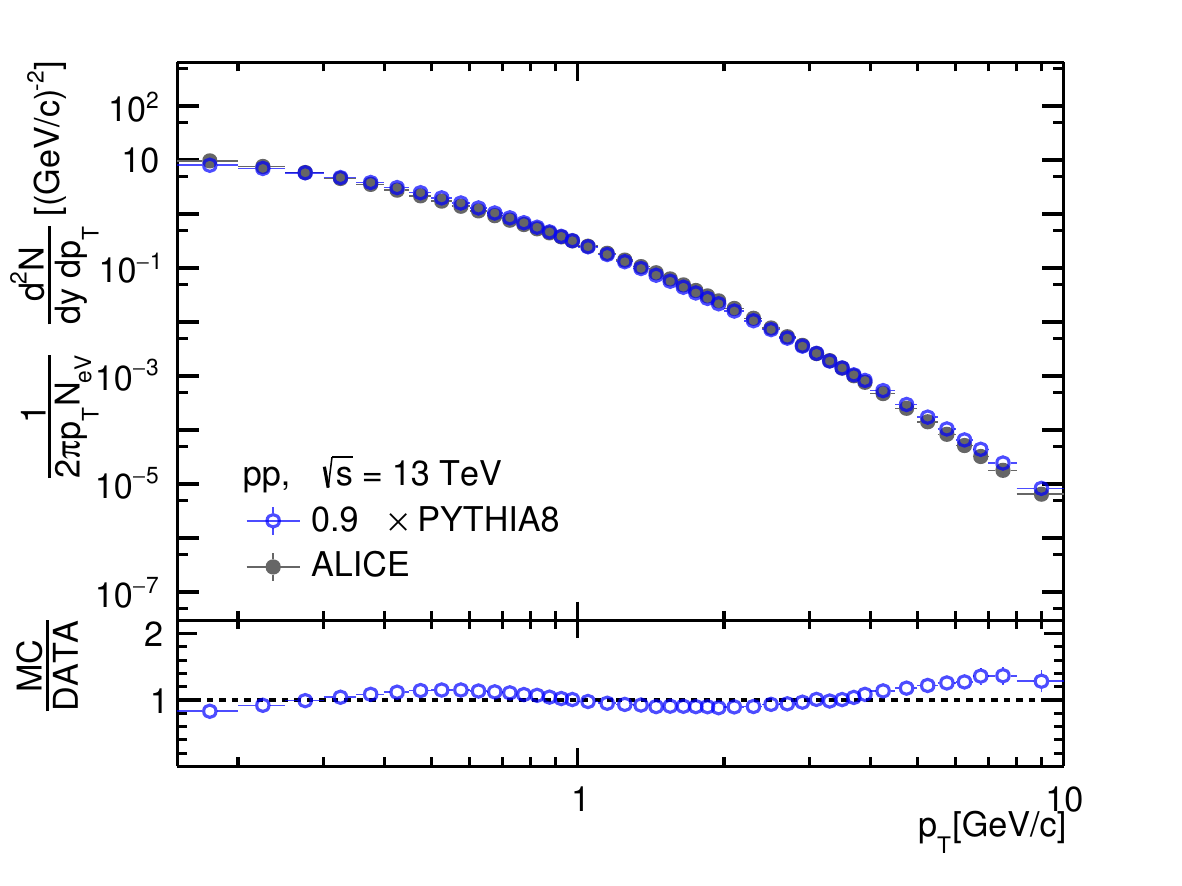}
    \caption{Transverse momentum spectra for minimum bias pp collisions at $\sqrt{s}=13$ TeV using PYTHIA~8 Color Ropes compared with the experimental measurements from ALICE~\cite{ALICE:2015qqj}.}
    \label{fig:pTAllCharged}
\end{figure}

In this study, we have used PYTHIA~8.308, an advanced version of PYTHIA~6, which incorporates the multi-partonic interactions scenario as one of the key improvements. This analysis is performed by generating $6\times 10^{7}$ events in pp collisions at  $\sqrt{s}$ = 13 TeV with Monash 2013 Tune (Tune:pp = 14)~\cite{Skands:2014pea}. We contemplate inelastic, soft QCD simulated events (SoftQCD:inelastic=on), which allows all the single, double, and central diffractive components of the total scattering cross-section. We have considered mode 1 of color reconnection (ColourReconnection:mode = 1) along with MPI (PartonLevel:MPI = on). This newer CR scheme is combined with mode 1 of beam remnants (BeamRemnants:remnantMode = 1). Mode 1 of color reconnection is a QCD-based model which minimizes the string length and the color rules from QCD. As a result, the QCD multiplets can produce triplets and junctions, which produce more baryons. Furthermore, along with CR, we employ another mechanism of hadronization for color strings called rope hadronization. The tunings of the rope hadronization used in this study are similar to the string shoving mechanism introduced in PYTHIA~8. The details of the parameters used for rope hadronization are given in Table~\ref{table:rope}. To set impact-parameter-plane vertices for partonic production by MPI, FSR, ISR, and beam remnants, we use the flag partonvertex (PartonVertex:setVertex = on). For the generated events, we let all the resonances decay except the ones used in our study with the switch HadronLevel:Decay = on. The particles originating from the MPIs and the beam remnants form the underlying events. The Lund-string fragmentation model performs the hadronisation of these partons \cite{Andersson:1983ia}. The CR picture ensures that the string between the partons is arranged in such a way that the total string length is reduced, which in turn leads to reduced particle multiplicity of the event. From here onwards in the paper, the mention of Color Ropes means PYTHIA~8 Monash 2013 tune with the inclusion of QCD-based color reconnection and rope hadronization mechanisms. On the other hand, Monash implies the PYTHIA~8 default 2013 Monash tune, and Monash NoCR corresponds to PYTHIA~8 2013 Monash without color reconnection.

\begin{figure}
    \centering
    \includegraphics[scale=0.4]{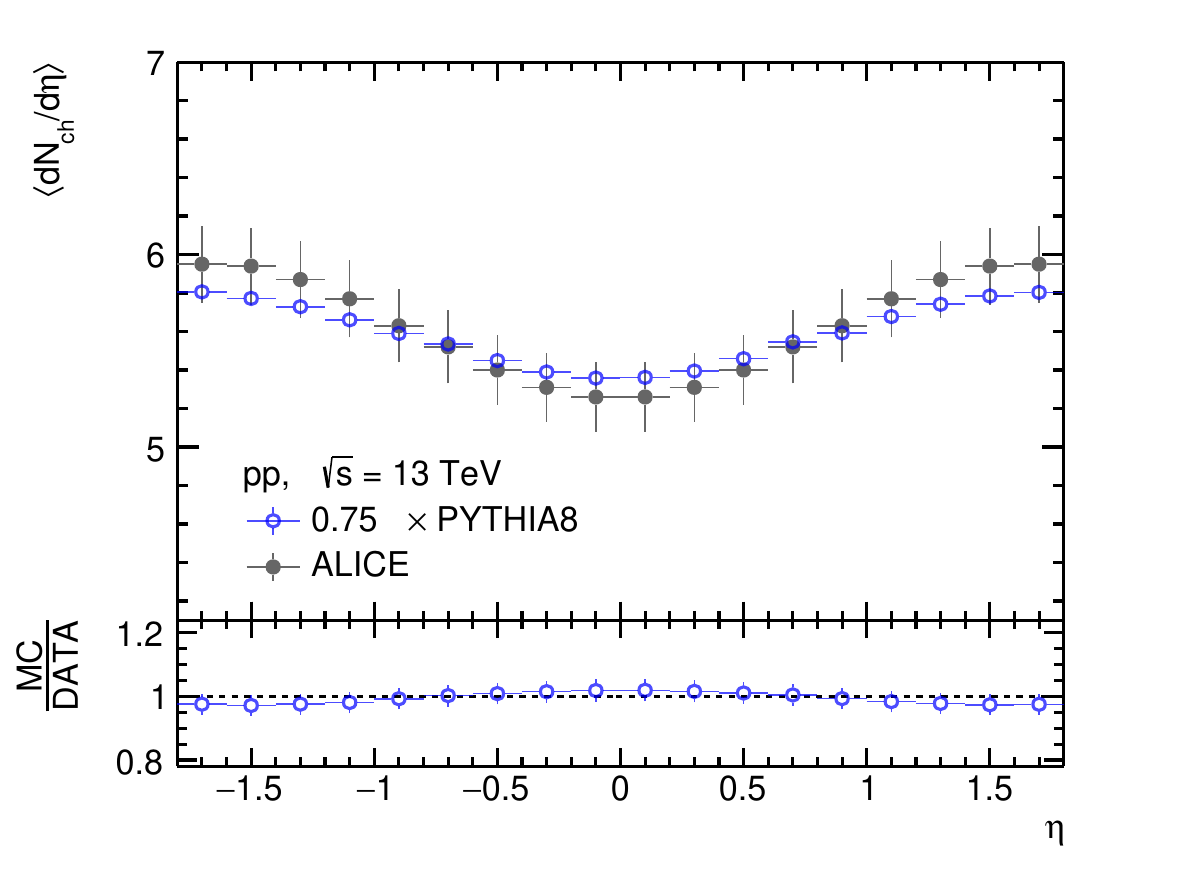}
    \caption{Psudorapidity distribution for minimum bias pp collisions at $\sqrt{s}=13$ TeV using PYTHIA~8 Color Ropes compared with the experimental measurements from ALICE~\cite{ALICE:2015qqj}.}
    \label{fig:EtaAllCharged}
\end{figure}

To check the compatibility of PYTHIA~8 Color Ropes with experimental data, we have compared the production cross section obtained from PYTHIA~8 as a function of transverse momentum and pseudorapidity with the ALICE experimental data~\cite{ALICE:2015qqj} in pp collisions at $\sqrt{s}$ = 13 TeV for all charged particles and shown in Figs.~\ref{fig:pTAllCharged} and ~\ref{fig:EtaAllCharged}, respectively. As can be seen from the bottom panels, the spectral shape prediction from PYTHIA~8 is consistent with experimental data within ~5\%.

\subsection{Event classifiers at the LHC}
In this subsection, we define different event shape observables used at the LHC recently.

\subsubsection{Number of multi-partonic interactions ($N_{\rm mpi}$)}

At LHC and RHIC, the hadronic collisions create a large number of particles in the final state. The basic understanding of particle production in such hadronic collisions can be made by QCD improved parton model~\cite{Sjostrand:2004pf}. Here, the hadrons are described as the composite of elementary particles, such as quarks and gluons, held together by strong force. This composite nature of hadrons leads to the occurrence of events with multiple partonic scatterings or MPI in minimum-bias hadronic collisions. In addition to initial hard scatterings, most inelastic processes in hadronic collisions can contain several perturbatively calculable QCD interactions when one extends the pQCD theory to low-$p_{\rm T}$ with some distance above $\Lambda_{\rm QCD}$~\cite{Sjostrand:2004pf}. These perturbatively calculable QCD interactions, called semi-hard processes and contribute to MPI significantly, are of much importance as they can affect the final state multiplicity~\cite{Sjostrand:2004pf}. Thus, with an increase in the number of semi-hard processes, the number of multi-partonic interactions ($N_{\rm mpi}$) rises. The events with a large $N_{\rm mpi}$ tend to result in large final state multiplicity and vice versa. Thus, the study of $N_{\rm mpi}$ is crucial for understanding particle production in the final state in elementary hadronic collisions.

\subsubsection{Charged particle multiplicity ($N_{\rm ch}$)}

The charged particle multiplicity estimator is used to study the multiplicity dependence of particle production dynamics. It is considered as one of the most common event classifier at the LHC energies. In the ALICE experiment at the LHC for INEL $>$ 0 sample the events are classified, and the charged particle multiplicity is measured based on the total charge deposited in the V0 detector (V0M amplitude) and on the number of silicon pixel detector (SPD) tracklets ($N_{\rm  SPD tracklets}$) in the pseudorapidity region $|\eta| < 0.8$. However, the charged particle multiplicity also can be measured at two different pseudorapidity regions. The V0 detector with acceptance of $-3.7 < \eta < -1.7$ and $2.8 < \eta < 5.1$  measures the charged particle multiplicity at the forward pseudorapidity region and is denoted as $N_{\rm ch}^{\rm fwd}$. On the other hand, the SPD and the Time Projection Chamber (TPC) measure the charged particle multiplicity at the mid-pseudorapidity region (\textit{i.e.}, $|\eta| < 0.8$) and is denoted as $N_{\rm ch}^{\rm mid}$. Studying the particle production dynamics with different multiplicity estimators allows us to understand potential biases that arise due to the multiplicity measurement in the overlapping $\eta$ regions.
 

\subsubsection{Transverse sphericity ($S_{\rm{T}}$)}

Another interesting set of event shape observables that characterizes the underlying event (UE) shape based on the momentum of the charged hadrons is estimated from the linearized sphericity tensor~\cite{Hanson:1975fe, CMS:2018mdd, ALICE:2012cor};

\begin{equation}
    S^{\rm \mu \nu}= \frac{\sum_{i =1}^{N_{\rm ch}} p^{{\rm \mu}}_{i} p^{{\rm \nu}}_{i}/|p_{i}| }{\sum_{i =1}^{N_{\rm ch}} |p_{i}|}
    \label{Smunu}
\end{equation}
where the index $i$ runs over all the charged hadrons associated with a given event, and $\mu$ and $\nu$ indices refer to one of the (x, y, z) components of the momentum of a particle. The momentum tensor $S^{\rm \mu \nu}$ has three eigenvalues $\lambda_1, \lambda_2,  \lambda_3$. The individual eigenvalues are normalized and ordered such that $\lambda_1 > \lambda_2 > \lambda_3$ with $\lambda_1+\lambda_2+\lambda_3$ =1 by definition. Using these eigenvalues, one can construct various event shape observables such that;

\begin{itemize}
    \item Aplanarity: A = $\frac{3}{2}$$ \lambda_3$ measures the amount of transverse momentum in or out of the plane formed by the two leading order eigenvectors and can be measured by the smallest eigenvalue $\lambda_3$.  The allowed values of A lie between $0 \leq A < 1/2$; however, the typical measured values lie between $0 \leq A < 0.3$.  The value of $A$ close to zero indicates the relatively planar events, while the value of $A$ close to 1/2 indicates the isotropic events.

    \item Sphericity: S = $\frac{3}{2} (\lambda_2 + \lambda_3) $ measures the total transverse momentum with respect to the sphericity axis defined by the four momenta used for the event shape measurement (usually, the first eigenvector). The allowed values lie between $0 \leq S < 1$, but due to the inclusion of the smallest eigenvalue $ \lambda_3$, the typical maximum value achieved in the experiment is around S $\simeq 0.8$. The value of S close to zero indicates a dijet kind of event, while the value of S close to 1 indicates the isotropic events.

    \item C = 3($\lambda_1 \lambda_2 + \lambda_1 \lambda_3 + \lambda_2 \lambda_3$)
    measures the events with 3 jets. The value of C tends to 0 for dijet events.

    \item D = 27$\lambda_1 \lambda_2  \lambda_3 $
    measures the events with 4 jets. The value of D tends to 0 for dijet or trijet events.
    
\end{itemize}
 In hadron colliders, the event shape analysis is mostly restricted to the transverse plane to avoid the biases coming from the boost along the beam axis~\cite{ALICE:2012cor}. Therefore, similar to sphericity as defined above, another well-known event shape classifier is introduced in the transverse plane called transverse sphericity ($S_{\rm{T}}$). It is one of the widely used event shape observables at Stanford Linear Accelerator Center (SLAC) to identify the existence of jets in leptonic collisions with collision energies up to 7.4 GeV~\cite{Hanson:1975fe}. Following Eq.~\ref{Smunu}, the transverse sphericity can be defined in terms of the eigenvalues of the transverse momentum matrix ($S_{\rm xy}^{\rm Q}$), given 
 as~\cite{ALICE:2012cor}:
\begin{equation}
    S_{\rm xy}^{\rm Q}=\frac{1}{\sum_{i}p_{{\rm T}_{i}}}\sum_{i}\begin{pmatrix}
p_{{\rm x}_{i}}^{2} & p_{{\rm x}_{i}}p_{{\rm y}_{i}} \\
p_{{\rm y}_{i}}p_{{\rm x}_{i}} & p_{{\rm y}_{i}}^{2}
\end{pmatrix} \text{[GeV/$c$].}
\label{SQxy}
\end{equation}
Here, $(p_{{\rm x}_{i}},p_{{\rm y}_{i}})$ are the projections of transverse momentum ($p_{\rm T_{i}}$) of $i$th particle in $(x,y)$ directions. The quadratic nature of $S_{\rm xy}^{\rm Q}$ in the particle momenta makes the $S_{\rm xy}^{\rm Q}$ matrix to be a non-collinear safe quantity in pQCD. To make it collinear-safe, Eq.~\ref{SQxy} can be linearized as follows,
\begin{equation}
    S_{\rm xy}^{\rm L}=\frac{1}{\sum_{i}p_{{\rm T}_{i}}}\sum_{i}\frac{1}{p_{{\rm T}_{i}}}\begin{pmatrix}
p_{{\rm x}_{i}}^{2} & p_{{\rm x}_{i}}p_{{\rm y}_{i}} \\
p_{{\rm y}_{i}}p_{{\rm x}_{i}} & p_{{\rm y}_{i}}^{2}
\end{pmatrix} 
\label{SLxy}
\end{equation}

Similar to sphericity, the transverse sphericity can be defined in terms of the eigenvalues of $S_{\rm xy}^{\rm Q}$ matrix, i.e., $\lambda_{1}$, and $\lambda_{2}$, with $\lambda_1>\lambda_2$, as follows.

\begin{eqnarray}
    S_{\rm T}=\frac{2\lambda_{2}}{\lambda_{1}+\lambda_{2}}
    \label{ST}
\end{eqnarray}

By construction, $S_{\rm T}$ ranges between 0 and 1, where $S_{\rm T}\rightarrow 0$ denotes the events are jetty or pencil-like, while $S_{\rm T}\rightarrow 1$ signifies the isotropic events dominated with the soft production of particles. Transverse sphericity, in this work, has been estimated in the mid-pseudorapidity region, i.e., $|\eta|<0.8$ with charged hadrons having $p_{\rm T}>0.15$ GeV/$c$. For the calculation of $S_{\rm T}$, only the events with $N_{\rm ch}^{\rm mid}\ge 10$ are considered.

\subsubsection{Transverse spherocity ($S_{0}$)}

 Based on the different geometrical distributions of final state particles produced in hadronic and nuclear collisions, another event shape observable is constructed called transverse spherocity, which separates the events based on their geometrical shapes. Transverse spherocity is defined as follows~\cite{Cuautle:2014yda, Cuautle:2015kra, Ortiz:2017jho, Banfi:2010xy, Khuntia:2018qox}
\begin{equation}
    S_{0} = \frac{\pi^{2}}{4}\min_{\hat{n}} \bigg(\frac{\sum_{i=1}^{N_{\rm had}}|\vec p_{{\rm T}_{i}}\times\hat{n}|}{\sum_{i=1}^{N_{\rm had}}~p_{{\rm T}_{i}}}\bigg)^{2}
\label{eq:sphero}
\end{equation}

Here, the unit vector $\hat{n} (n_{T},0)$ is chosen in such a way that it minimizes ratio $S_{0}$ presented within the parentheses of Eq.~\ref{eq:sphero}. It is usually calculated using charged particle tracks in the mid-pseudorapidity ($|\eta|<0.8$) that have $p_{\rm T} > $ 0.15 GeV/$c$. In this study, $S_{0}$ is calculated with more than 10 charged particle tracks with $|\eta|<0.8$ and $p_{\rm T}>0.15$ GeV/c in an event to ensure a statistically meaningful concept of topology. In Eq.~\ref{eq:sphero}, $N_{\rm had}$ refers to the total number of charged hadrons, and the multiplication factor of $\pi^2/4$ ensures that the $S_{0}$ estimator varies from 0 to 1. The two extreme limits of spherocity correspond to the two different configurations of event topology. The value $S_{0} \rightarrow$ 0 corresponds to events with single back-to-back jets, while $S_{0} \rightarrow$ 1 corresponds to events with isotropic emission of particle production.  Conventionally, the events located in the bottom 20\% of the spherocity distribution are referred to as jetty events, while the top 20\% of the spherocity distribution are referred to as isotropic events.

It is important to note that Eq.~\ref{eq:sphero} is the $p_{\rm T}$ weighted definition of $S_{0}$ estimator. Experimentally, this weighted $S_{0}$ estimator introduces a neutral jet bias and detector smearing effect as discussed in Ref~\cite{ALICE:2023bga}. This issue could be fixed by modifying the definition via a new estimator called the unweighted transverse spherocity estimator ($S_{0}^{p_{\rm T} = 1}$). This is the unweighted transverse spherocity estimator by setting $p_{\rm T}=1$ for all the charged particles in Eq.~\ref{eq:sphero}. Hence, the unweighted transverse spherocity ($S_{0}^{p_{\rm T} = 1}$) is defined as follows

\begin{equation}
    S_{0}^{p_{\rm T} = 1}=\frac{\pi^2}{4}\min_{\hat{n}}\Bigg(\frac{\sum_{i=1}^{N_{\rm had}}|\hat{p_{\rm T}} \times \hat{n}|}{N_{\rm had}}\Bigg)^{2}.
    \label{eq:spheropt1}
\end{equation}

Since the neutral jet biases were towards high-$p_{\rm T}$ particles, in Eq.~\ref{eq:spheropt1} it is removed by setting $p_{\rm T}=1$. Furthermore, the denominator of Eq.~\ref{eq:sphero}, i.e., $\sum_{i}p_{{\rm T}_i}$ is now replaced with the number of charged hadrons, i.e., $N_{\rm had}$. It has been observed that with the unweighted transverse spherocity estimator, the detector smearing effect is smaller than the weighted transverse spherocity estimator~\cite{ALICE:2023bga}.

It is noteworthy that the transverse spherocity is less likely to reach the isotropic limit than sphericity~\cite{Banfi:2010xy,Ortiz:2017jho}. Therefore, the discrimination power between isotropic soft events and symmetric multi-jet events might be the highest for spherocity.

\subsubsection{Relative transverse activity classifier ($R_{\rm T}$)}

In recent days, a new event activity estimator has been proposed called the relative transverse activity classifier ($R_{\rm T}$) to understand the selection bias introduced towards the hard processes, when one selects pp collisions with high event activity. This selection bias may affect the measurement and observable of interest. It is proposed that these biases can be minimized by removing the jet contribution from the event activity estimator, which can be achieved by identifying an axis that allows for the event-by-event separation of the jet contribution from the UE. In this study, the direction of the leading charged particle with the highest transverse momentum (trigger particle) is used as a reference axis to build the particle correlation with the associated particles in azimuthal angle $\Delta\phi$ = $\phi^{trig} - \phi^{asso}$. \\

Thus, to separate the events based on the contribution of jets from the UE, different topological regions are defined namely called towards, transverse, and away regions. The towards region consists of the tracks with an azimuthal angle less than $\pi/3$ relative to the trigger particle, \textit{i.e.}, $|\Delta\phi|<\pi/3$, while the away region contains all the tracks with relative azimuthal angle $|\Delta\phi|\geq 3\pi/2$. Furthermore, the transverse region is defined to be the relative azimuthal angle within $\pi/3\leq|\Delta\phi|<3\pi/2$. The towards and the away regions have the largest contribution from the fragmentation of jets originating from the hardest partonic interactions in an event and are expected to be insensitive to the softer UE. However, the transverse region is perpendicular to the leading jet axis, it is expected to have the least contribution from the jet and must be dominated by the UE activity, thus a better region to build the event activity classifier $R_{\rm T}$. \\

Therefore, to estimate the value of $R_{\rm T}$, one needs the number of charged hadrons in the transverse regions ($N_{\rm ch}^{\rm T}$) per event and the average charged particle multiplicity analyzed in the whole sample (${\langle N_{\rm ch}^{\rm T}\rangle}$). Mathematically, the value of $R_{\rm T}$ is defined as follows;

\begin{equation}
    R_{\rm T}=\frac{N_{\rm ch}^{\rm T}}{\langle N_{\rm ch}^{\rm T}\rangle}
    \label{eq:RT}
\end{equation}




In Eq.~\ref{eq:RT}, the angular bracket represents the event-average value of the observables. $R_{\rm T}$ is measured using the charged hadrons at $|\eta|<0.8$. For the estimation of $R_{\rm T}$, only the events having leading $p_{\rm T}\ge 5$ GeV/c at $|\eta|<0.8$ are considered. Here, a large value of $R_{\rm T}$ implies events dominated with the underlying event activity in contrast to a smaller value of $R_{\rm T}$.

\subsubsection{Charged particle flattenicity ($\rho_{\rm{ch}}$)}

For isotropic events, a near uniform distribution of transverse momentum is expected throughout the $\eta-\phi$ phase space. To measure the uniformity of transverse momentum distribution event-by-event, the $\eta-\phi$ region can be divided into $(10\times 10)$ grid. Thus, by measuring the transverse momentum in each cell ($p_{\rm T}^{\rm cell}$), one can define the charged particle flattenicity ($\rho_{\rm{ch}}$) as follows~\cite{Ortiz:2022zqr},
\begin{equation}
    \rho_{\rm{ch}}=\frac{\sigma_{p_{\rm T}^{\rm cell}}}{\langle p_{\rm T}^{\rm cell}\rangle}
    \label{eq:flatpt}
\end{equation}

where, $\langle p_{\rm T}^{\rm cell}\rangle$ denotes the mean  and $\sigma_{p_{\rm T}^{\rm cell}}$ standard deviation of the transverse momentum distribution.

Furthermore, Eq.~\ref{eq:flatpt}, can be written as
\begin{equation}
    \rho_{\rm{ch}}=\frac{\sqrt{\sum_{i}(p_{\rm T}^{\rm{cell}, i}-\langle p_{\rm T}^{\rm{cell}}\rangle)^2/N_{\rm{cell}}}}{\langle p_{\rm T}^{\rm cell}\rangle}
    \label{eq:flatpt2}
\end{equation}

Keeping the ALICE 3 tracking and particle identification (PID) capabilities in mind, the charged particles in $|\eta|<4.0$ and $p_{\rm T}>0.15$ GeV/c are considered for the estimation of $\langle p_{\rm T}^{\rm cell}\rangle$ and $\sigma_{p_{\rm T}^{\rm cell}}$. It is expected that events with jet signals along with underlying event activity have a larger spread along with the higher $\sigma_{p_{\rm T}^{\rm cell}}$. On the other hand, events having only soft production of particles will have lower spread in $p_{\rm T}^{\rm cell}$, and corresponding $\sigma_{p_{\rm T}^{\rm cell}}$ will be lower. Thus, the value of $\rho_{\rm{ch}}$ is expected to be smaller for isotropic events and large for jetty events.

However, the current detector capabilities for PID and tracking at ALICE at the LHC and STAR at the RHIC experiments are limited to central rapidity region $|\eta|<1.0$. Thus, using the above definition of charged particle flattenicity defined in Eq.~\ref{eq:flatpt2}, one can not estimate the value of $\rho_{\rm{ch}}$ at the experiments. Furthermore, the measurement of both the event shape and the particle of interest in the same pseudorapidity region creates a bias. This bias can be minimized by choosing the event shape estimator in the forward rapidity region, which is in the different pseudorapidity region compared to the particle of interest~\cite{ALICE:2018pal}. Most of the present detectors can measure the charged particle multiplicity in the forward pseudorapidity. 
Therefore, to measure the charged particle flattenicity with the current detector scenarios, a redefinition of flattenicity is required, which is redefined in Ref.~\cite{Ortiz:2022mfv}.  According to the new definition of charged particle flattenicity, $\rho_{\rm ch}$ can be estimated using the following equation.
\begin{equation}
    \rho_{\rm ch} = \frac{\sqrt{\sum_{i}(N_{\rm ch}^{\rm{cell}, i}-\langle N_{\rm ch}^{\rm{cell}}\rangle)^2/N^{2}_{\rm{cell}}}}{\langle N_{\rm ch}^{\rm{cell}}\rangle}
    \label{eq:flatNch}
\end{equation}

In this definition, the charged particle multiplicity is measured at the forward rapidity instead of the transverse momentum of the particles. To calculate $\rho_{\rm ch}$, the ($\eta-\phi$) phase space is divided into ($8\times 8$) cells.  The average charged particle multiplicity in each cell '$i$' ($N_{\rm ch}^{\rm{cell}, i}$) and the average of $N_{\rm ch}^{\rm{cell}, i}$ ($\langle N_{\rm ch}^{\rm{cell}}\rangle$) in an event is required to evalute $\rho_{\rm ch}$. Similar to other standard event activity, by construction, $\rho_{\rm ch}$ ranges from 0 to 1~\cite{Ortiz:2022mfv}.  The lower limit of 0 indicates isotropic events, while 1 indicates jetty events. To be consistent with other event shape observables (e.g, sphericity, spherocity, etc.), here, throughout the manuscript, we use $1-\rho_{\rm ch}$ instead of $\rho_{\rm ch}$.  The events with $1-\rho_{\rm ch}\rightarrow 1$ are likely to be isotropic while events with $1-\rho_{\rm ch}\rightarrow 0$ exhibit jetty topology.

\begin{figure}[ht!]
\begin{center}
\includegraphics[scale=0.4]{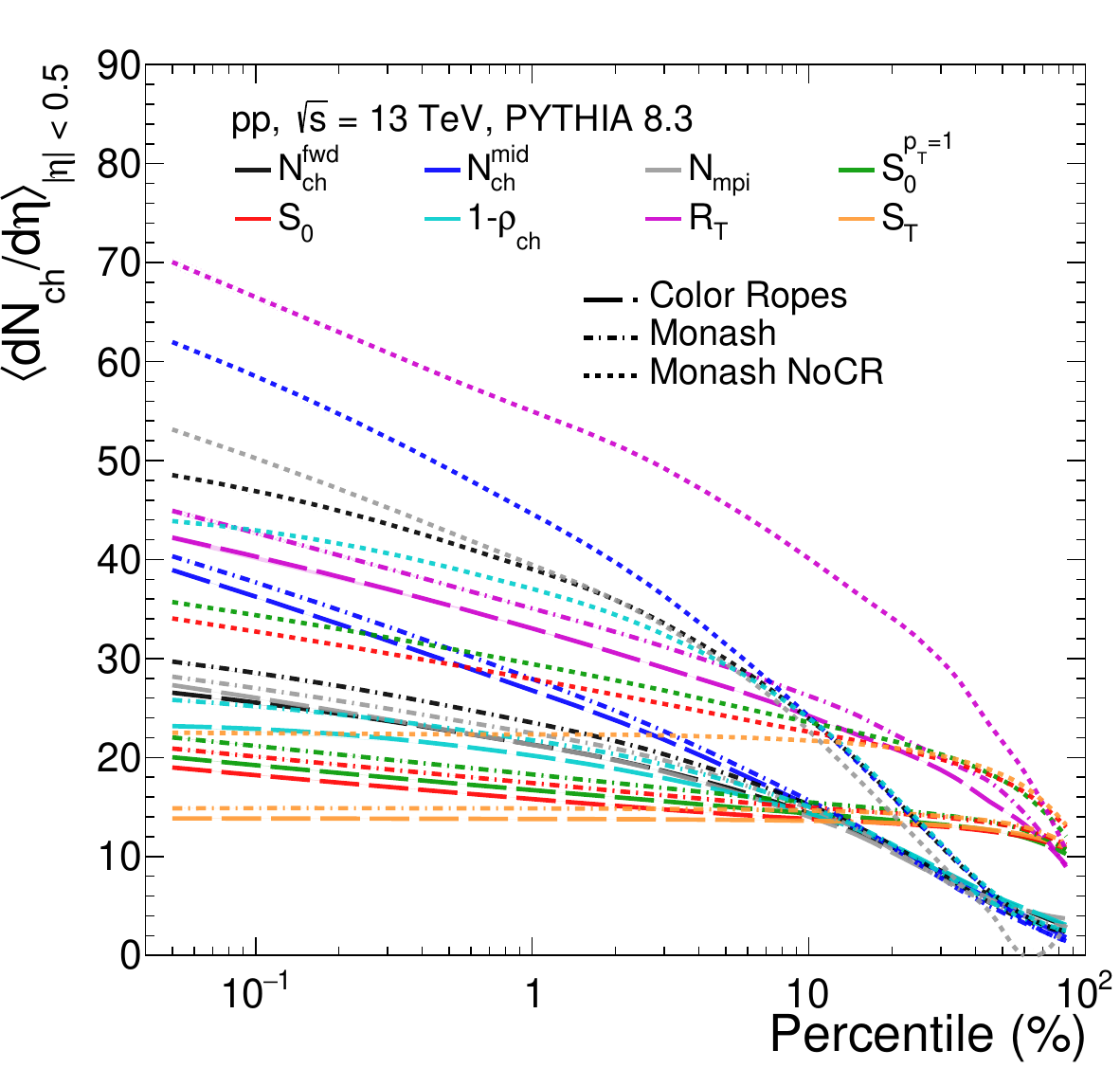}
\caption{Mapping of different percentiles of event classifiers with the charged particle multiplicity at mid-pseudorapidity in pp collisions at $\sqrt{s}=13$ TeV using PYTHIA~8.}
\label{fig:EsvsMult}
\end{center}
\end{figure}

Table~\ref{tab:ESCuts} shows the selection cuts used in this study for different event classifiers in pp collisions at $\sqrt{s}=13$ TeV using PYTHIA~8. The corresponding mapping of different percentiles of event shapes with charged particle multiplicity at mid-pseudorapidity regions for PYTHIA~8 Color Ropes can be found in Table~\ref{tab:dnchdeta}. In Fig.~\ref{fig:EsvsMult}, we show the correlation of event classifier percentiles with charged particle density at the midrapidity ($\langle dN_{\rm ch}/d\eta\rangle_{|\eta|<0.5}$) in pp collisions at $\sqrt{s}=13$ TeV using Color Ropes, Monash and Monash NoCR tunes of PYTHIA~8. As one moves from lower to higher percentiles of event classes, a decrease in $\langle dN_{\rm ch}/d\eta\rangle_{|\eta|<0.5}$ is observed. However, the rate of decrease of $\langle dN_{\rm ch}/d\eta\rangle_{|\eta|<0.5}$ from lower to higher percentiles of event classifiers is lower for $S_{0}$, $S_{0}^{p_{\rm T}=1}$ and $S_{\rm T}$ for all considered cases of PYTHIA~8 as compared to other event classifiers. Further, one finds that the Monash NoCR has the largest values of $\langle dN_{\rm ch}/d\eta\rangle_{|\eta|<0.5}$ in the lowest percentiles of all the event classifiers. In contrast, Color Ropes possess the smallest values. Since, without CR, the individual partonic interactions are independent, leading to a large final state multiplicity. With CR in the picture, each parton from individual partonic interaction can color reconnect to form hadrons, which have larger $\langle p_{\rm T}\rangle$ while reduced particle multiplicity in the final state~\cite{OrtizVelasquez:2013ofg}. Further, with color ropes in the picture, some of the energy is expended to enhance the production of strange hadrons, which further reduces $\langle dN_{\rm ch}/d\eta\rangle_{|\eta|<0.5}$. It is to be noted that, due to the event selection cut of 10 charged tracks in $|\eta|<0.8$, $\langle dN_{\rm ch}/d\eta\rangle_{|\eta|<0.5}$ is larger in the higher percentiles for $S_0$, $S_0^{p_{\rm T}=1}$, and $S_{\rm T}$ as compared to other event classifiers. Further, the requirement of events with leading $p_{\rm T}>5$ GeV/c leads to large multiplicity for $R_{\rm T}$ in the final state.

\begin{table*}[]
    \caption{ Selection cuts in terms of percentiles for different event classifiers in pp collisions at $\sqrt{s}=13$ TeV using PYTHIA~8 Color Ropes.}
    \centering
    \begin{tabular}{|c|c|c|c|c|c|c|c|c|}
    \hline
         Percentile & $N_{\rm mpi}$ & $N_{\rm ch}^{\rm mid}$ & $N_{\rm ch}^{\rm fwd}$ &  $S_{\rm T}$ & $S_{0}$ &  $S_{0}^{p_{\rm T}=1}$  & $N^{\rm T}_{\rm ch}$ & 1-$\rho_{\rm ch}$  \\  \hline \hline
         0 - 0.1 & 23 - 35 & 54 - 150 & 120 - 250 & 0.98 - 1&  0.92 - 1 & 0.94 - 1 &   26 - 60 & 0.91 - 1 \\
         \hline
         0.1 - 1 & 18 - 23 & 40 - 54 & 96 - 120 & 0.96 - 0.98 &  0.88 - 0.92 & 0.91 - 0.94  & 20 - 26 & 0.90 - 0.91   \\
         \hline
         1 - 5 & 14 - 18 & 29 - 40 & 74 - 96 & 0.91 - 0.96 & 0.82 - 0.88 & 0.87 - 0.91 & 15 - 20 & 0.88 - 0.90 \\
         \hline
         5 - 10 & 11 - 14 & 23 - 29 & 62 - 74 & 0.88 - 0.91 & 0.78 - 0.82 & 0.84 - 0.87 & 13 - 15 & 0.87 - 0.88 \\
         \hline
         10 - 20 & 8 - 11 & 17 - 23 & 47 - 62 & 0.82 - 0.88 & 0.72 - 0.78 & 0.79 - 0.84 & 10 - 13 & 0.85 - 0.87 \\
         \hline
         20 - 30 & 5 - 8 & 13 - 17 & 37 - 47 & 0.78 - 0.82 & 0.67 - 0.72 & 0.75 - 0.79 & 9 - 10 & 0.83 - 0.85\\
         \hline
         30 - 40 & 4 - 5 & 10 - 13 & 30 - 37 & 0.73 - 0.78 & 0.62 - 0.67 & 0.72 - 0.75 & 7 - 9 & 0.81 - 0.83\\
         \hline
         40 - 50 & 3 - 4 & 8 - 10 & 24 - 30 & 0.68 - 0.73 & 0.56 - 0.62 & 0.68 - 0.72 & 6 - 7 & 0.79 - 0.81\\
         \hline
         50 - 70 & 2 - 3 & 5 - 8 & 16 - 24 & 0.59 - 0.68 & 0.46 - 0.56 & 0.59 - 0.68 & 4 - 6 & 0.74 - 0.79\\
         \hline
         70 - 100 & 0 - 2 & 1 - 5 & 1 - 16 & 0 - 0.59 & 0 - 0.46 & 0 - 0.59 & 0 - 4 & 0 - 0.74 \\
         \hline
    \end{tabular}
    \label{tab:ESCuts}
\end{table*}

\begin{table*}[]
    \caption{ The averaged charged particle multiplicity density calculated in the mid-rapidity ($|\eta| < 0.5$) for each event classifiers in pp collisions at $\sqrt{s}=13$ TeV using PYTHIA~8 Color Ropes.}
    \centering
    \begin{tabular}{|c|c|c|c|c|c|c|c|c|}
    \hline
         Percentile & $N_{\rm mpi}$ & $N_{\rm ch}^{\rm mid}$ & $N_{\rm ch}^{\rm fwd}$ &  $S_{\rm T}$ & $S_{0}$ &  $S_{0}^{p_{\rm T}=1}$  & $N^{\rm T}_{\rm ch}$ & 1-$\rho_{\rm ch}$  \\  \hline \hline
         0 - 0.1 & 27.31 $\pm$ 0.03 & 38.94 $\pm$ 0.02 & 26.55 $\pm$ 0.03 & 13.85 $\pm$ 0.02 & 18.99 $\pm$ 0.03  & 20.03 $\pm$ 0.03 & 42.22 $\pm$ 0.28    & 23.17 $\pm$ 0.03 \\
         \hline
         0.1 - 1 & 22.54 $\pm$ 0.01 & 29.29 $\pm$ 0.01& 22.52 $\pm$ 0.01 & 13.81 $\pm$ 0.01 & 16.41 $\pm$ 0.01   & 17.33 $\pm$ 0.01 & 35.07 $\pm$ 0.07 &21.15 $\pm$ 0.01  \\
         \hline
         1 - 5 & 18.55 $\pm$ 0.0 & 21.80 $\pm$ 0.0 & 18.63 $\pm$ 0.0 & 13.75 $\pm$ 0.01 & 14.75 $\pm$ 0.01   & 15.58 $\pm$ 0.01 & 25.06 $\pm$ 0.03 & 17.92 $\pm$ 0.01 \\
         \hline
         5 - 10 & 15.21 $\pm$ 0.0 & 16.79 $\pm$ 0.0 & 15.54 $\pm$ 0.0 & 13.65 $\pm$ 0.01 & 13.98 $\pm$ 0.01   & 14.62 $\pm$ 0.01 & 25.51 $\pm$ 0.03 & 15.35 $\pm$ 0.0 \\
         \hline
         10 - 20 & 12.5 $\pm$ 0.0 & 12.84 $\pm$ 0.0 & 12.5 $\pm$ 0.0 & 13.5 $\pm$ 0.0 & 13.54 $\pm$ 0.0   & 13.96 $\pm$ 0.0 & 22.28 $\pm$ 0.02 & 12.71 $\pm$ 0.0   \\
         \hline
         20 - 30 & 9.01 $\pm$ 0.0 & 9.63 $\pm$ 0.0 & 9.66 $\pm$ 0.0 & 13.27 $\pm$ 0.0 & 13.18 $\pm$ 0.0   & 13.44 $\pm$ 0.0 & 19.84 $\pm$ 0.02 & 9.84 $\pm$ 0.0   \\
         \hline
         30 - 40 & 6.98 $\pm$ 0.0 & 7.34 $\pm$ 0.0 & 7.65 $\pm$ 0.0 & 13.02 $\pm$ 0.0 & 12.88 $\pm$ 0.0   & 13.05 $\pm$ 0.0 & 17.68 $\pm$ 0.01 & 7.71 $\pm$ 0.0  \\
         \hline
         40 - 50 &  5.84 $\pm$ 0.0 & 5.71 $\pm$ 0.0 & 6.13 $\pm$ 0.0 & 12.70 $\pm$ 0.0 & 12.54 $\pm$ 0.0   & 12.67 $\pm$ 0.0 & 15.54 $\pm$ 0.02 & 6.22 $\pm$ 0.0 \\
         \hline
         50 - 70 & 4.59 $\pm$ 0.0 & 4.03 $\pm$ 0.0 & 4.63 $\pm$ 0.0 & 12.2 $\pm$ 0.0 & 12.01 $\pm$ 0.0   & 11.94 $\pm$ 0.0 & 13.27 $\pm$ 0.01 & 4.79 $\pm$ 0.0 \\
         \hline
         70 - 100 &  3.72 $\pm$ 0.0 & 1.82 $\pm$ 0.0 & 2.99 $\pm$ 0.0 & 10.91 $\pm$ 0.0 & 10.82 $\pm$ 0.0   & 10.26 $\pm$ 0.0 & 9.15 $\pm$ 0.01 & 3.08 $\pm$ 0.0 \\
         \hline
    \end{tabular}
    \label{tab:dnchdeta}
\end{table*}

\section{Results and discussions}
\label{results}

\begin{figure*}[ht!]
\begin{center}
\includegraphics[scale=0.29]{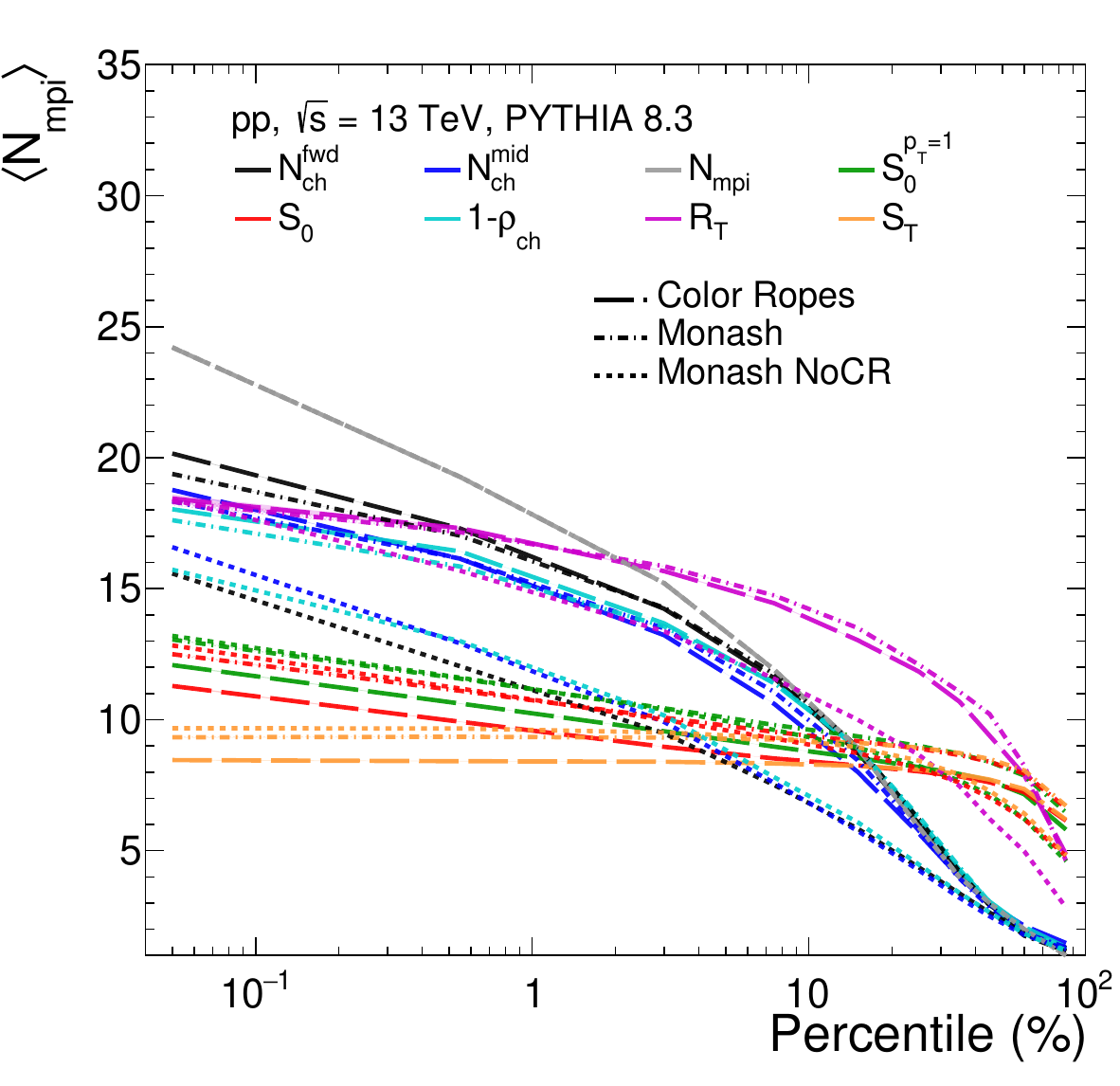}
\includegraphics[scale=0.29]{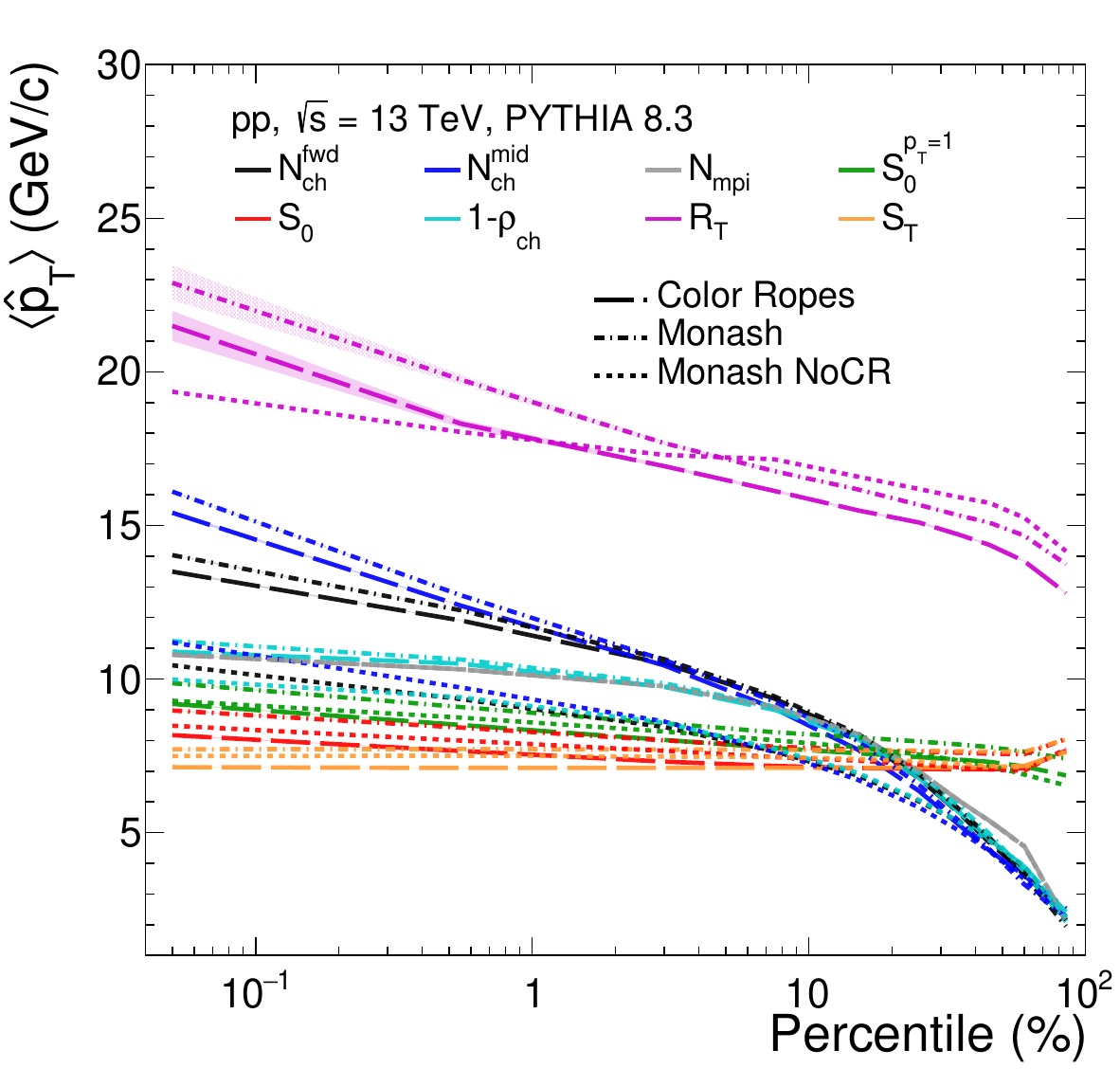}
\includegraphics[scale=0.29]{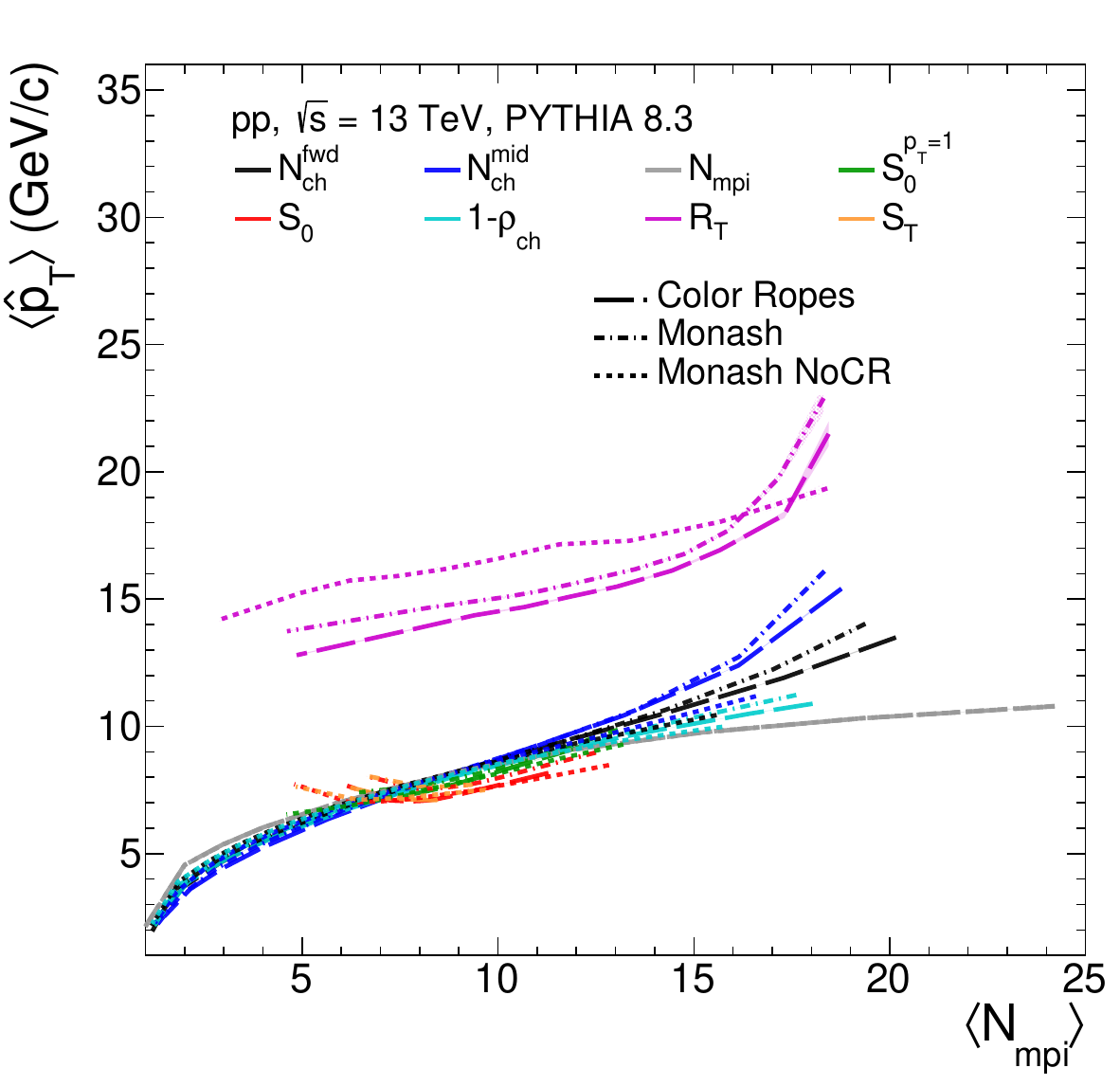}
\caption{Mapping of different percentiles of event classifiers with $\langle N_{\rm mpi}\rangle$ (left) and $\langle \hat{p}_{\rm T}\rangle$ (middle) in pp collisions at $\sqrt{s}=13$ TeV using PYTHIA~8. The right panel shows the correlation between $\langle N_{\rm mpi}\rangle$ and $\langle \hat{p}_{\rm T}\rangle$ as a function of different percentiles of event classifiers. Comparisons are made for Color Ropes, Monash and Monash NoCR tunes of PYTHIA~8.}
\label{fig:MPIvspThat}
\end{center}
\end{figure*}

The RH mechanism in PYTHIA~8 with the events having large partonic interactions can lead to enhanced production of strange hadrons. Thus, it becomes crucial to understand the correlation of different event classifiers used in this study with the $N_{\rm mpi}$ for different modes of PYTHIA~8. Further, certain event classifiers can be prone to selection biases coming from different sources, which should also be properly investigated to understand the microscopic origin of strangeness enhancement probed by different event classifiers. Thus, in Fig.~\ref{fig:MPIvspThat}, we show the correlation of different event classifiers with $N_{\rm mpi}$ and transverse momentum transfer of the hardest parton–parton interaction ($\hat{p}_{\rm T}$). The left panel of Fig.~\ref{fig:MPIvspThat} shows the correlation of event shape percentiles with the average number of multi-partonic interactions ($\langle N_{\rm mpi}\rangle$) in pp collisions at $\sqrt{s}=13$ TeV using PYTHIA~8. The comparisons with Color Ropes, Monash and Monash NoCR tunes of PYTHIA~8 are also shown. In the left panel of Fig.~\ref{fig:MPIvspThat}, we observe that the lower percentiles of event classifiers probe to a higher value of $\langle N_{\rm mpi}\rangle$ as compared to higher percentiles. In PYTHIA~8 simulation with Color Ropes, one finds that $N_{\rm ch}^{\rm fwd}$ probes to a higher values of $\langle N_{\rm mpi}\rangle$, followed by $N_{\rm ch}^{\rm mid}$, $R_{\rm T}$ and $1-\rho_{\rm ch}$ which probes a similar region of $\langle N_{\rm mpi}\rangle$ in the lowest percentiles. In contrast, event classifiers, such as $S_{0}^{p_{\rm T}=1}$, $S_{0}$, and $S_{\rm T}$ have the lowest correlation with $\langle N_{\rm mpi}\rangle$ in the order $\langle N_{\rm mpi}\rangle_{S_{0}^{p_{\rm T}=1}} > \langle N_{\rm mpi}\rangle_{S_{0}} > \langle N_{\rm mpi}\rangle_{S_{\rm T}}$. This is also true for PYTHIA 8 Monash events, except here $\langle N_{\rm mpi}\rangle_{R_{\rm T}} \approx \langle N_{\rm mpi}\rangle_{N_{\rm ch}^{\rm mid}} > \langle N_{\rm mpi}\rangle_{1-\rho_{\rm ch}}$. In contrast, for the PYTHIA~8 Monash NoCR case, one finds a slightly different order of event classifiers that probe to a higher value of $\langle N_{\rm mpi}\rangle$ in the lowest percentiles. Here, the order is $\langle N_{\rm mpi}\rangle_{R_{\rm T}} > \langle N_{\rm mpi}\rangle_{N_{\rm ch}^{\rm mid}} > \langle N_{\rm mpi}\rangle_{1-\rho_{\rm ch}} \approx \langle N_{\rm mpi}\rangle_{N_{\rm ch}^{\rm fwd}} > \langle N_{\rm mpi}\rangle_{S_{0}^{p_{\rm T}=1}} > \langle N_{\rm mpi}\rangle_{S_{0}} > \langle N_{\rm mpi}\rangle_{S_{\rm T}}$. Further, the event selections with $N_{\rm ch}^{\rm fwd}$, $N_{\rm ch}^{\rm mid}$, $R_{\rm T}$ and $1-\rho_{\rm ch}$ are able to probe to higher $\langle N_{\rm mpi}\rangle$ for the Color Ropes case, while the sensitivity to $\langle N_{\rm mpi}\rangle$ is minimum for Monash NoCR, except for $R_{\rm T}$, which shows a similar $\langle N_{\rm mpi}\rangle$ at the lowest percentile. In contrast, $S_{0}^{p_{\rm T}=1}$, $S_{0}$, and $S_{\rm T}$ are found to have reduced sensitivity to $\langle N_{\rm mpi}\rangle$ for events with PYTHIA~8 Color Ropes as compared to Monash and Monash NoCR. In addition, from the left panel of Fig.~\ref{fig:MPIvspThat}, it is evident that depending upon different hadronisation mechanisms, the values of different event shape classifiers and their sensitivity to $\langle N_{\rm mpi}\rangle$ can vary, where some observables are found to better at events with Color Ropes while others are more effective in PYTHIA~8 Monash NoCR to probe intrinsic $\langle N_{\rm mpi}\rangle$.

The middle panel of Fig.~\ref{fig:MPIvspThat} shows the variation of 
average transverse momentum transfer of the hardest parton–parton interaction ($\langle\hat{p}_{\rm T}\rangle$) as a function of percentiles of different event classifiers in pp collisions at $\sqrt{s}=13$ TeV using Color Ropes, Monash and Monash NoCR tunes of PYTHIA~8. Here, as one moves from lower to higher percentile values of event classifiers, one finds a decrease in $\langle \hat{p}_{\rm T}\rangle$. The values of $\langle \hat{p}_{\rm T}\rangle$ towards the lowest percentiles of event classifiers, such as $N_{\rm ch}^{\rm mid}$, $N_{\rm ch}^{\rm fwd}$, $R_{\rm T}$ and $1-\rho_{\rm ch}$ having a strong correlation with $\langle dN_{\rm ch}/d\eta\rangle_{|\eta|<0.5}$, shown in Fig.~\ref{fig:EsvsMult}, are smaller for Monash NoCR as compared to Color Ropes and Monash cases. This is evident as a large multiplicity in the Monash NoCR case would lead to less energetic hadrons formed in the collisions. Additionally, $\langle\hat{p}_{\rm T}\rangle$  is found to be the largest for $R_{\rm T}$ for all the considered cases of PYTHIA~8 because the estimation of $R_{\rm T}$ requires events having at least one charged particle with $p_{\rm T}> 5$ GeV/$c$, which biases the sample towards events having jets. Interestingly, when events are selected with $N_{\rm mpi}$, we find a saturating behaviour in $\langle \hat{p}_{\rm T}\rangle$ towards the lower percentiles, which is closely reproduced by event selection with $1-\rho_{\rm ch}$ for the events generated with Color Ropes and Monash tunes of PYTHIA~8. In addition, the events selected with charged particle multiplicities, such as $N_{\rm ch}^{\rm mid}$ and $N_{\rm ch}^{\rm fwd}$ show a continuous rise in $\langle \hat{p}_{\rm T}\rangle$ with decrease in percentile even towards the lowest percentiles, which indicates that high multiplicity events can have significant contributions from multi-jet topologies~\cite{ALICE:2022qxg, ALICE:2023plt}. Further, $S_{0}^{p_{\rm T}=1}$, $S_{0}$, and $S_{\rm T}$ have minute sensitivity to $\langle \hat{p}_{\rm T}\rangle$ as compared to other event classifiers. 

With the correlation study of different event classifiers with $\langle N_{\rm mpi}\rangle$ and $\langle \hat{p}_{\rm T}\rangle$, now we dig for event shape observables that closely follow the event selection based on $N_{\rm mpi}$ baseline. Therefore, we study the correlation between $\langle N_{\rm mpi}\rangle$ and $\langle \hat{p}_{\rm T}\rangle$ as shown in the right panel of Fig.~\ref{fig:MPIvspThat} for different percentiles of event classifiers with Color Ropes, Monash and Monash NoCR tunes of PYTHIA~8. With increase in $\langle N_{\rm mpi}\rangle$ corresponding $\langle \hat{p}_{\rm T}\rangle$ rises for all classifiers. However, as discussed earlier, $R_{\rm T}$ shows enhanced values of $\langle \hat{p}_{\rm T}\rangle$ as the events are biased towards the jets. Further, the rise of $\langle \hat{p}_{\rm T}\rangle$ with $\langle N_{\rm mpi}\rangle$ is stronger for event selected with $N_{\rm ch}^{\rm mid}$ and $N_{\rm ch}^{\rm fwd}$ where contributions from multi-jet topologies are inferred. $S_{0}^{p_{\rm T}=1}$, $S_{0}$, and $S_{\rm T}$ probe a very limited region in $\langle \hat{p}_{\rm T}\rangle$ and $\langle N_{\rm mpi}\rangle$. Finally, events selected with $1-\rho_{\rm ch}$ closely follow $N_{\rm mpi}$.

So far, we understand the correlation of different event classifiers with $N_{\rm mpi}$ for different modes of PYTHIA~8 and possible biases present in the study. To have a more comprehensive understanding of the possible correlation among different event classifiers and the biases on particle production dynamics,  we explore the event shape observable dependence of primary strange ($K_{S}^{0}, \Lambda, \bar{\Lambda}$) and multi-strange ($\Xi^{+},\Xi^{-}, \Omega^{+}, \Omega^{-}$) hadrons production in pp collisions at $\sqrt{s}$ = 13 TeV using different tunes of PYTHIA~8. As discussed in the previous sections, multiplicity-based estimators are prone to selection bias coming from different sources. This motivates us to study the strange to non-strange ratio as a function of other event classifiers, such as the number of multi-partonic interactions, transverse sphericity, transverse spherocity, relative transverse activity classifier, and charged particle flattenicity. Here, we, mainly investigate $2K_{S}^{0}/(\pi^{+}+\pi^{-})$, $(\Lambda+\bar{\Lambda})/ (\pi^{+}+\pi^{-})$, ($\Xi^{+} + \Xi^{-})/(\pi^{+}+\pi^{-})$, $2\phi/(\pi^{+}+\pi^{-})$, and $(\Omega^{+}+ \Omega^{-})/(\pi^{+}+\pi^{-})$ ratios as a function of different event classifiers. For simplicity now onwards we refer these ratios as $K_{S}^{0}/\pi$, $\Lambda/\pi$, $ \Xi/\pi $, $\phi/\pi$, and $\Omega/\pi$ respectively. \\

 \begin{figure}[ht!]
\begin{center}
 \includegraphics[scale=0.43]{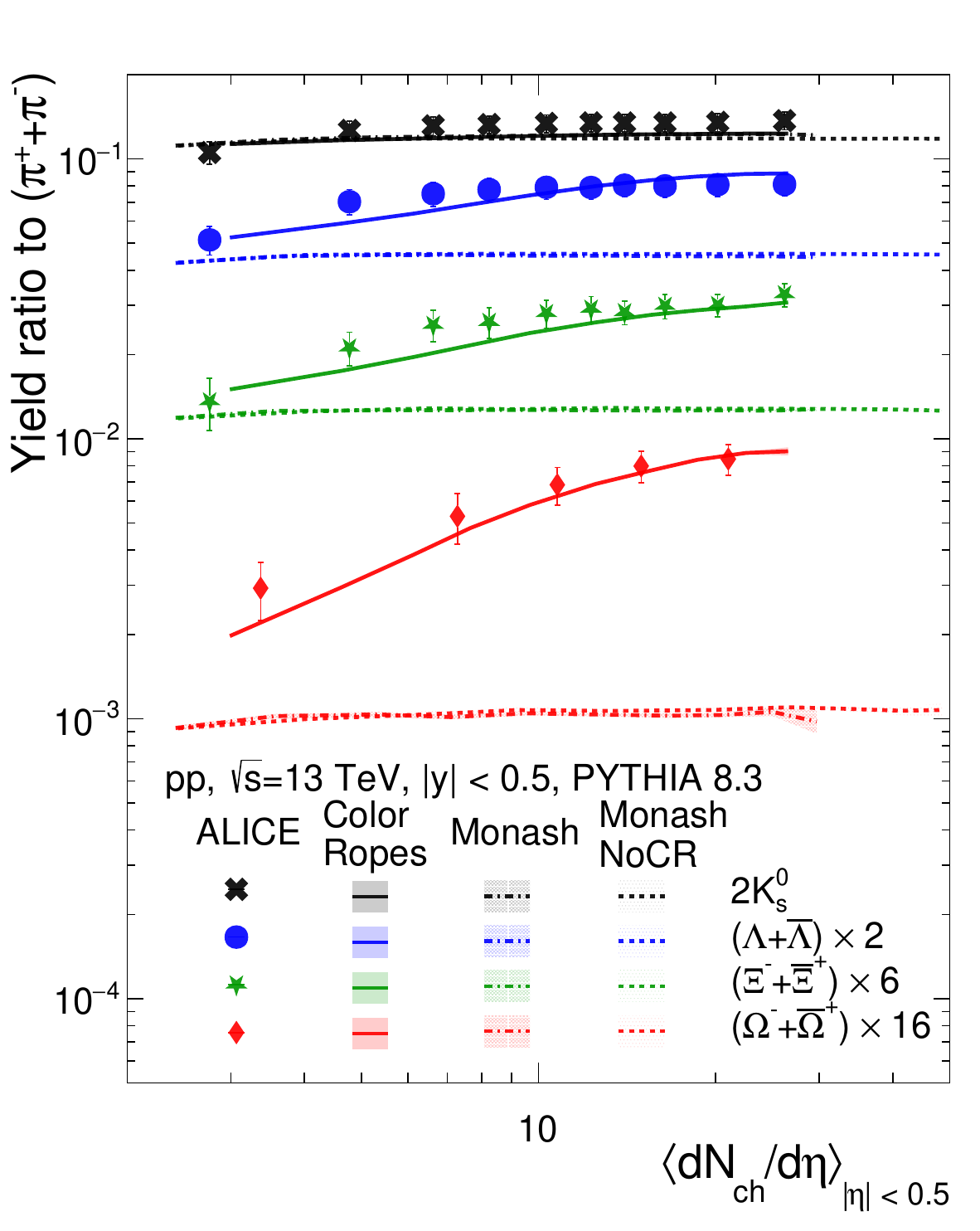}
\caption{$p_{\rm T}$-integrated yield ratios to pions ($\pi^{+}+\pi^{-}$) obtained in $|y|<0.5$ as a function of mean charged particle multiplicity ($\langle dN_{\rm ch}/d\eta\rangle$) obtained at $|\eta|<0.5$ in pp collisions at $\sqrt{s}=13$ TeV using different tunes of PYTHIA~8, and compared with similar measurements at ALICE~\cite{ALICE:2020nkc}.}
\label{fig:strangedatacomp}
\end{center}
\end{figure}


 \begin{figure}
\begin{center}
\includegraphics[scale=0.43]{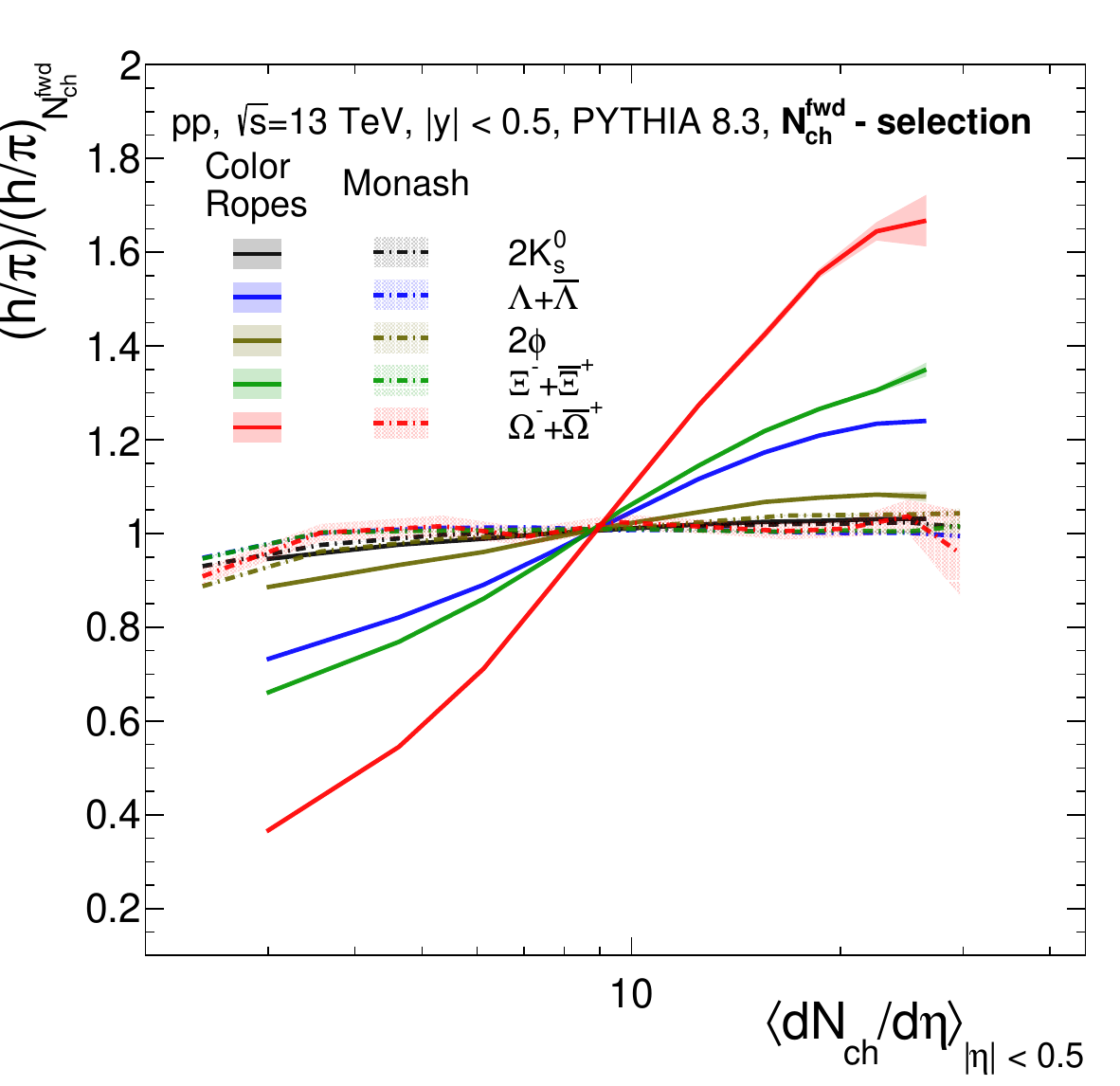}
\caption{Particle yield ratios to pions normalized to the values obtained for the minimum bias events as a function of $\langle dN_{\rm ch}/d\eta\rangle$ for different classes of $N_{\rm ch}^{\rm fwd}$  in pp collisions at $\sqrt{s}=13$ TeV using Color Ropes and Monash tunes of PYTHIA~8.}
\label{fig:SNYieldNch}
\end{center}
\end{figure}

 We begin by comparing the results of the PYTHIA~8 model for strange and multi-strange particle production yield ratio to charged pions with the ALICE results. Figure~\ref{fig:strangedatacomp} shows the $p_{\rm T}$-integrated yield ratios to pions ($\pi^{+}+\pi^{-}$) obtained in $|y|<0.5$ as a function of mean charged particle multiplicity ($\langle dN_{\rm ch}/d\eta\rangle$) obtained at $|\eta|<0.5$ in pp collisions at $\sqrt{s}=13$ TeV using PYTHIA~8 with Color Ropes, Monash and Monash NoCR tunes. Here, $\langle dN_{\rm ch}/d\eta\rangle$ is obtained for different event classes defined based on $N_{\rm ch}^{\rm fwd}$. From the figure, it is evident that in experiments, the strange particle yield ratio to pions increases with an increase in $\langle dN_{\rm ch}/d\eta\rangle$. The yield ratio becomes largely prominent for particles having larger valence strange quarks. Here, $K^{0}_{S}/\pi$ remains almost independent with an increase in charged particle multiplicity; however, the dependence grows when one moves from $\Lambda/\pi$ to $\Xi/\pi$ and $\Omega/\pi$, which have one, two and three valence strange quarks, respectively. The results from PYTHIA~8 with Color Ropes have both quantitative and qualitative agreement with the ALICE results. In contrast, Monash and Monash NoCR fail to explain the experimental data. The yield ratios of different strange hadrons to pions are similar for Monash and Monash NoCR tunes of PYTHIA~8 and significantly smaller as compared to results obtained from Color Ropes, except for $K^0_S$. Thus, henceforth, we shall be limiting our results to Color Ropes and Monash tunes of PYTHIA~8. Here, we do not observe any variation of self-normalised yield ratios of strange hadrons to pions a function of $\langle dN_{\rm ch}/d\eta\rangle$ for the Monash case.

 Figure~\ref{fig:SNYieldNch} shows the strange and multi-strange particle yield ratios to pions normalized to the value obtained for the minimum bias events as a function of $\langle dN_{\rm ch}/d\eta\rangle$ in pp collisions at $\sqrt{s}=13$ TeV using Color Ropes and Monash tunes of PYTHIA~8$^{\ref{note1}}$\footnotetext[1]{Note that we use $({\rm h}/\pi)_{\rm X}$ in the denominator of the $y$-axis of Fig. 6 and the lower panels of Figs.7-10, 12-14 to represent the hadron-to-pion ratio in $\rm X$-integrated events\label{note1}}. Here, the effect of strangeness enhancement in high multiplicity pp collisions with respect to the average minimum-bias events is visible for the events simulated with Color Ropes of PYTHIA~8, which is absent for the Monash case. Again, for the Color Ropes tune of PYTHIA~8, the hardons, which have a larger number of strange quarks, show an increasing trend with a higher slope, whereas, for $K^{0}_{S}$, the increasing slope is nearly zero. This observation is consistent with the experimental measurements with ALICE for pp collisions at $\sqrt{s}=7$ TeV~\cite{ALICE:2016fzo}. However, $\phi$ meson with zero net strange quantum number shows a stronger enhancement trend as compared to $K_{S}^{0}$ having one strange quark, which can make strangeness enhancement study through $\phi$ meson as a function of various event shapes interesting in small systems.
\begin{figure}
\begin{center}
\includegraphics[scale=0.4]{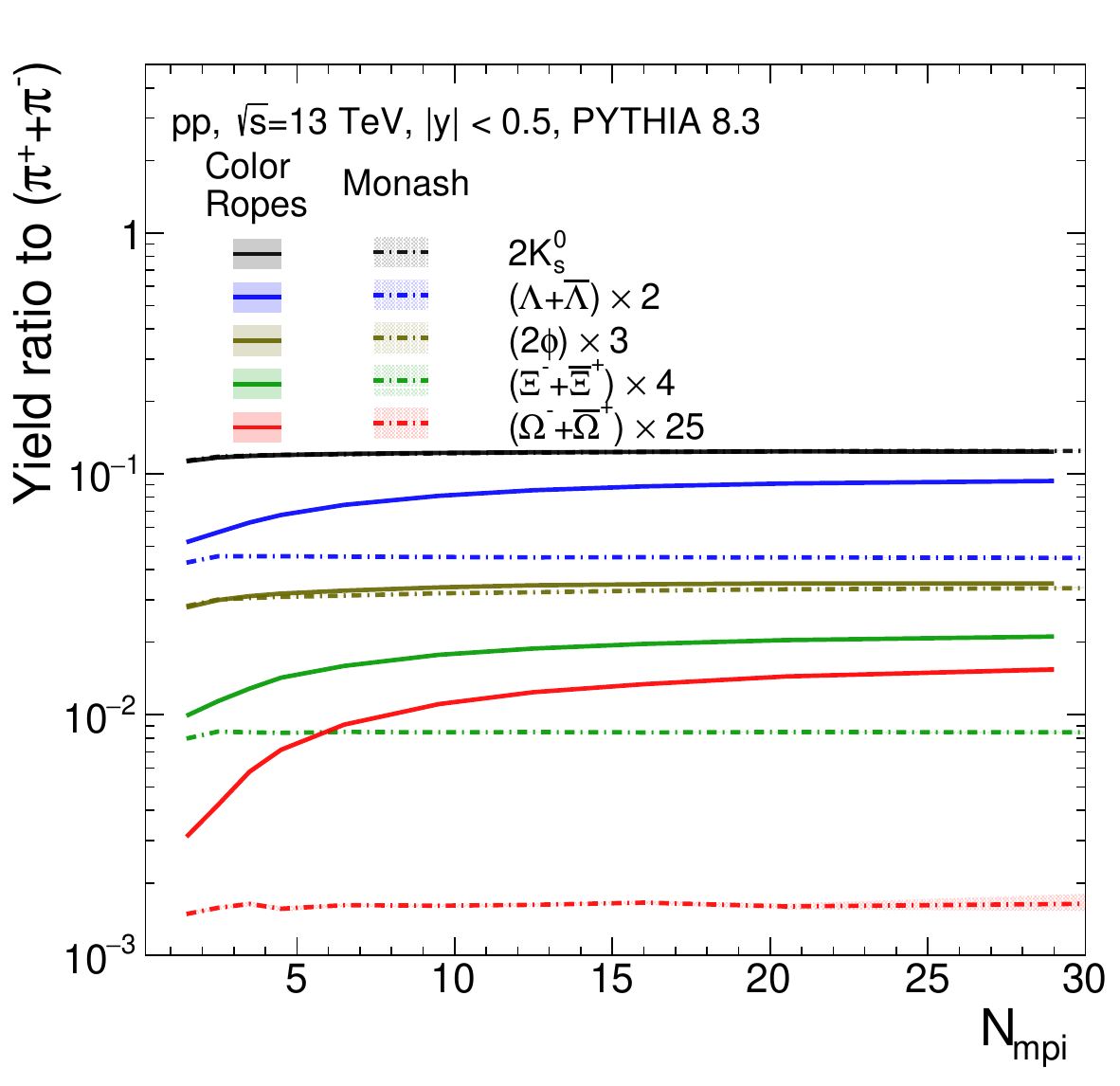}
\includegraphics[scale=0.4]{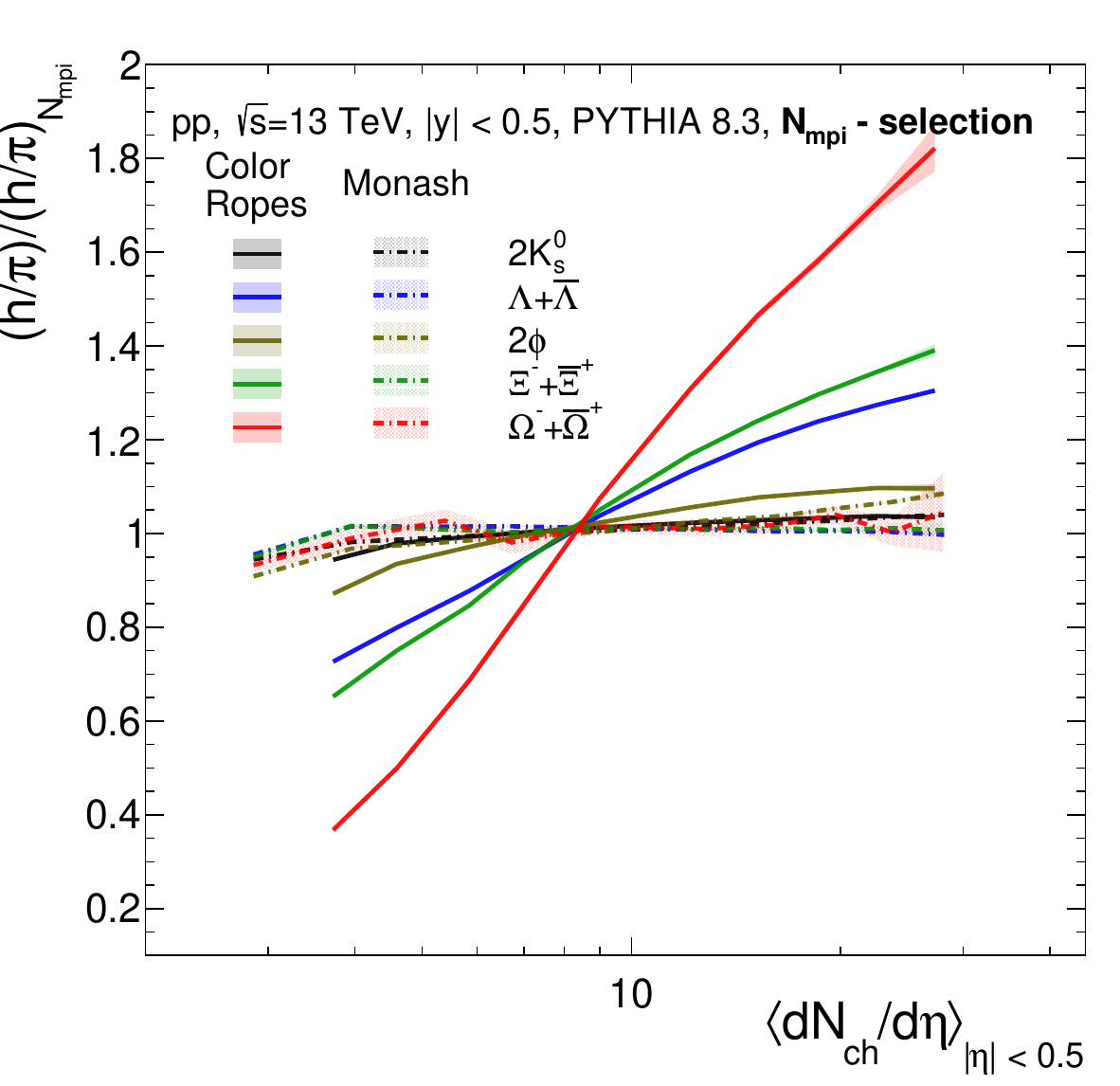}
\caption{Top: $p_{\rm T}$-integrated yield ratios to pions ($\pi^{+}+\pi^{-}$) obtained in $|y|<0.5$ as a function of number of multi-partonic interactions ($N_{\rm mpi}$) in pp collisions at $\sqrt{s}=13$ TeV using PYTHIA~8. Bottom: Particle yield ratios to pions normalized to the values obtained for the minimum bias events as a function of $\langle dN_{\rm ch}/d\eta\rangle$ in different classes of $N_{\rm mpi}$ in pp collisions at $\sqrt{s}=13$ TeV using Color Ropes and Monash tunes of PYTHIA~8.}
\label{fig:strangempi}
\end{center}
\end{figure}

After showing the feasibility of the PYTHIA~8 model, let us now study strange hadron production as a function of the number of multi-parton interactions ($N_{\rm mpi}$). The top panel of Fig.~\ref{fig:strangempi} shows the role of MPI in the production of strange hadrons having different numbers of valence strange quarks with respect to pions. Here, $p_{\rm T}$-integrated yield ratios of strange and multi-strange hadrons to pions are shown as a function of $N_{\rm mpi}$ in pp collisions at $\sqrt{s}=13$ TeV using PYTHIA~8. As one would expect, MPI plays a pivotal role in particle production, and it is also observed for strange and multi-strange hadrons. Here, the production of strange hadrons such as $K^{0}_{S}$, $\Lambda$, $\phi$, $\Xi$, and $\Omega$ particles increases with respect to that of pions as one goes from events having lower value of $N_{\rm mpi}$ to a large $N_{\rm mpi}$ events for the Color Ropes tune of PYTHIA~8. However, the yield ratios to pions seem to saturate towards higher values of $N_{\rm mpi}$. The effects of RH are further enhanced in the events having a large $N_{\rm mpi}$ value. A large value of $N_{\rm mpi}$ implies a larger number of color rope formations leading to enhanced production of strange and multi-strange baryons. However, $K^{0}_{S}/\pi$ and $\phi/\pi$ show negligible dependence on $N_{\rm mpi}$. Further, with an increase in $N_{\rm mpi}$ values, unlike Color Ropes tune of PYTHIA~8, the yield ratios of strange hadrons to pions obtained from Monash are nearly unchanged with $N_{\rm mpi}$.

In the bottom panel of Fig.~\ref{fig:strangempi}, we show the strange and multi-strange hadron yield ratios to pions scaled to the values obtained for the minimum bias events as a function of $\langle dN_{\rm ch}/d\eta\rangle$ for different classes of $N_{\rm mpi}$ in pp collisions at $\sqrt{s}=13$ TeV using PYTHIA~8. 
For the case with Color Ropes, the strange and multi-strange hadrons show a linear increase with the increase in $\log_{10}(\langle dN_{\rm ch}/d\eta\rangle)$ with events selected based on $N_{\rm mpi}$. The effects of RH are observed to be different for strange baryons ($\Lambda$, $\Xi$, and $\Omega$) from the strange meson ($K^{0}_{S}$) and hidden strange meson ($\phi$). Interestingly, $\Omega$, having three valence strange quarks, shows the largest enhancement with an increase in $N_{\rm mpi}$ in comparison to $\Lambda$, having only one valence strange quark. In contrast, the enhancement is negligible for $K_{s}^{0}$. Furthermore, the hidden strange meson, $\phi$, shows a small dependence on the event selection based on $N_{\rm mpi}$, similar to Fig.~\ref{fig:SNYieldNch}. In contrast, for the events generated with PYTHIA~8 Monash tunes, the self-normalised yield ratios to pions are found to be nearly flat with an increase in $\langle dN_{\rm ch}/d\eta\rangle$. This difference in the observation of different slopes of self-normalised yield ratios for Monash and Color Ropes cases of PYTHIA~8 indicates that MPI plays a fundamental and significant role in the production of strange hadrons when the RH mechanism is involved. 

Although, in experiments, the estimation of $N_{\rm mpi}$ is not possible, in this study, we try to probe the similar enhancement of different strange hadron production using different event shapes.

\begin{figure}
\begin{center}
\includegraphics[scale=0.4]{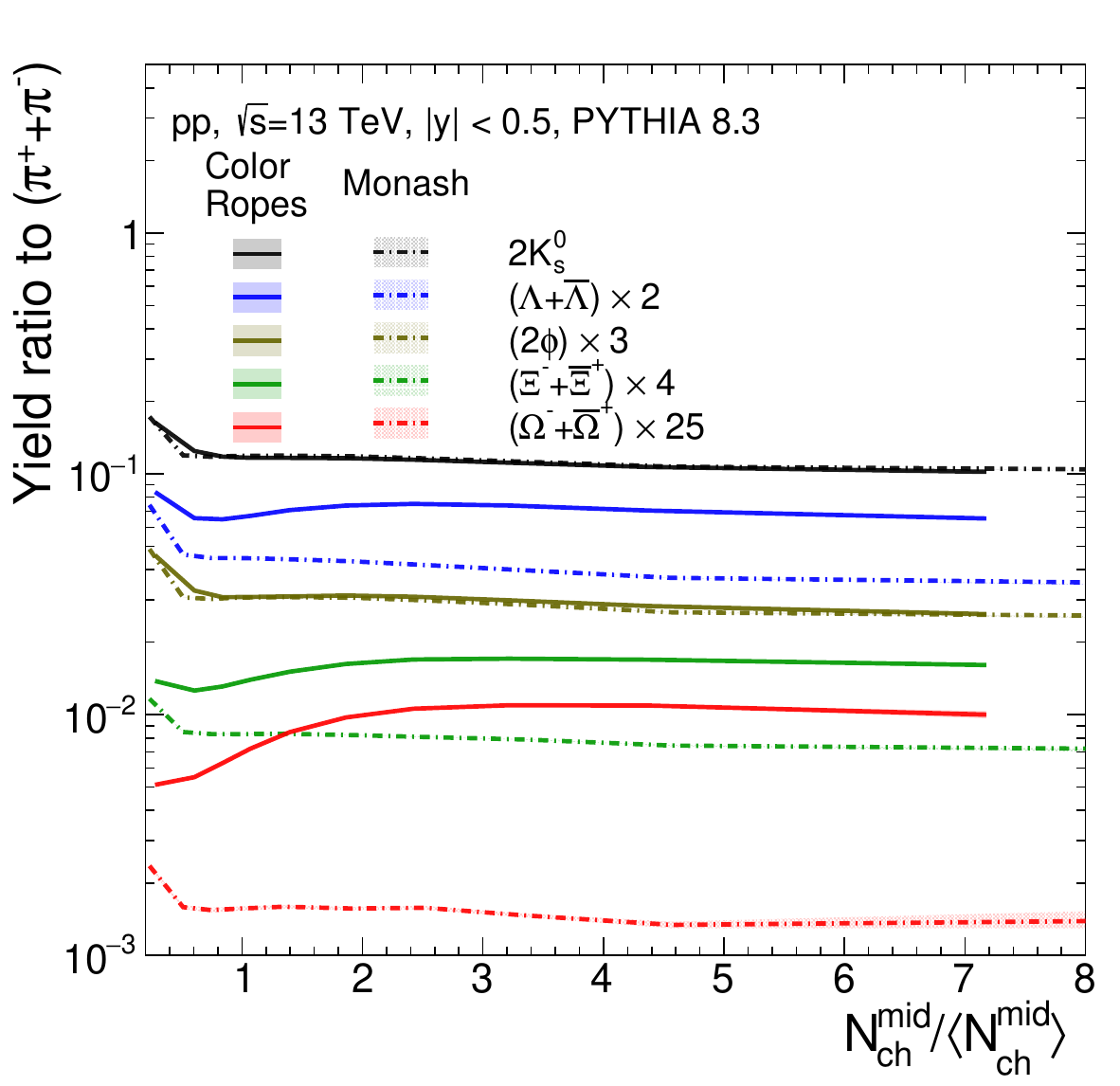}
\includegraphics[scale=0.4]{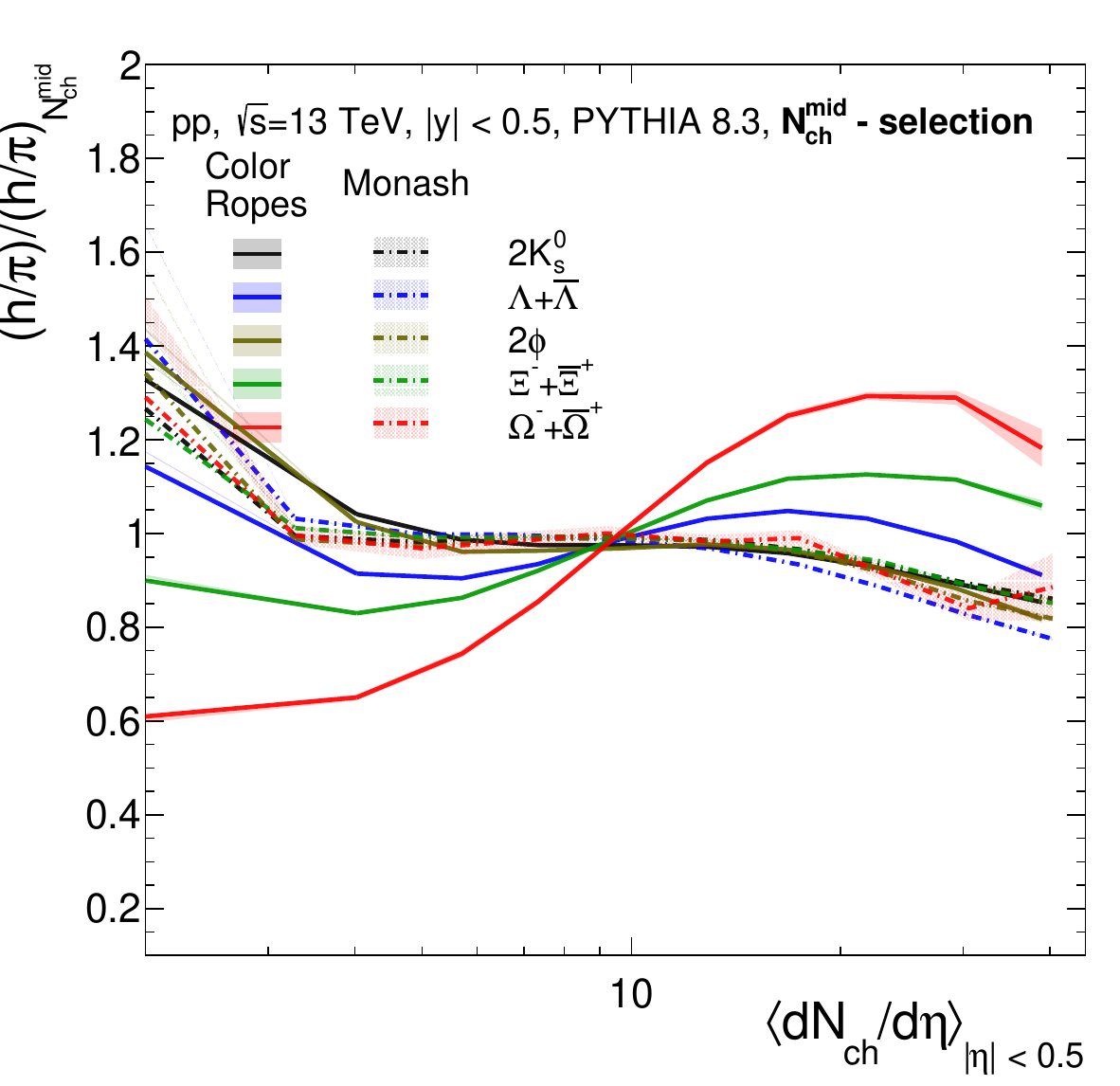}
\caption{$p_{\rm T}$-integrated yield ratios to pions ($\pi^{+}+\pi^{-}$) obtained in $|y|<0.5$ as a function of charged particle multiplicity obtained in the mid-pseudorapidity ($N_{\rm ch}^{\rm mid}$) (top) and forward pseudorapidity ($N_{\rm ch}^{\rm fwd}$) (bottom) in pp collisions at $\sqrt{s}=13$ TeV using Color Ropes and Monash tunes of PYTHIA~8.}
\label{fig:strangeNch}
\end{center}
\end{figure}


The top panel of Fig.~\ref{fig:strangeNch} shows the $p_{\rm T}$-integrated yield ratios of strange hadrons to pions as a function of self-normalised charged particle multiplicity obtained in the mid-pseudorapidity ($N_{\rm ch}^{\rm mid}/\langle N_{\rm ch}^{\rm mid}\rangle$) regions in pp collisions at $\sqrt{s}=13$ TeV using Color Ropes and Monash tunes of PYTHIA~8. Similar to Figs.~\ref{fig:strangedatacomp} and~\ref{fig:strangempi}, in Color Ropes case, the strangeness enhancement is found to be proportional to the strangeness content of the hadrons. For the Monash case, the strange particle yield ratios to pions show a negligible variation with $N_{\rm ch}^{\rm mid}/\langle N_{\rm ch}^{\rm mid}\rangle$.
The lower panel of Fig.~\ref{fig:strangeNch} shows the self-normalised yield ratios of strange hadrons to pions in pp collisions using Color Ropes and Monash tunes of PYTHIA~8. 
Here, a decrease in the strange to non-strange ratios in the high multiplicity pp collisions for both Color Ropes and Monash tunes of PYTHIA~8 is observed. This behaviour is not present when one studies the strangeness as a function of MPI and forward-rapidity multiplicity selection. This could be due to the auto-correlation bias introduced while selecting charged particle multiplicity in the mid-pseudorapidity region as the particle of interest is also studied in the mid-rapidity region~\cite{Weber:2018ddv}. This autocorrelation bias, where events are selected based on the charged particle multiplicity, which means mostly pions, causes the ratios $K^{0}_{S}/\pi$ and $\phi/\pi$ to decrease as the particle multiplicity increases. However, for $\Lambda$, the self-normalised yield ratio slightly increases in $7\lesssim\langle dN_{\rm ch}/d\eta\rangle\lesssim 15$, followed by a decrease in the lower and high $\langle dN_{\rm ch}/d\eta\rangle$ regions, which can be attributed to the competing effects of RH and autocorrelation bias due to event selection. Due to dominating contributions from RH, this effect is negligible for $\Xi/\pi$ and $\Omega/\pi$, which increase with increased charged particle multiplicity except for the highest multiplicity class. This can be further confirmed in the results of self-normalised yield ratios of strange hadrons to pions for the PYTHIA~8 Monash case. Here, due to the absence of RH, the self-normalised yield ratios of all the strange hadrons to pions decrease with an increase in charged particle multiplicity. In addition, due to this auto-correlation bias, the linear increment of the yield ratios scaled to the MB events with increasing $\log(\langle dN_{\rm ch}/d\eta\rangle)$ is absent when the events are selected based on the charged particle multiplicity at the midrapidity. Therefore, it can be inferred that the event selection in different charged particle multiplicity windows plays an important role in describing the dynamics of the strange hadron production. Furthermore, the relative production of the hidden strange $\phi$ particle slightly decreases with $N_{\rm ch}^{\rm mid}$, similar to the production of $K_{S}^{0}$. This implies that due to the autocorrelation bias of event selection, the neutral mesons behave similarly irrespective of their valence strange quark content.
 

\begin{figure}[ht!]
\begin{center}
\includegraphics[scale=0.4]{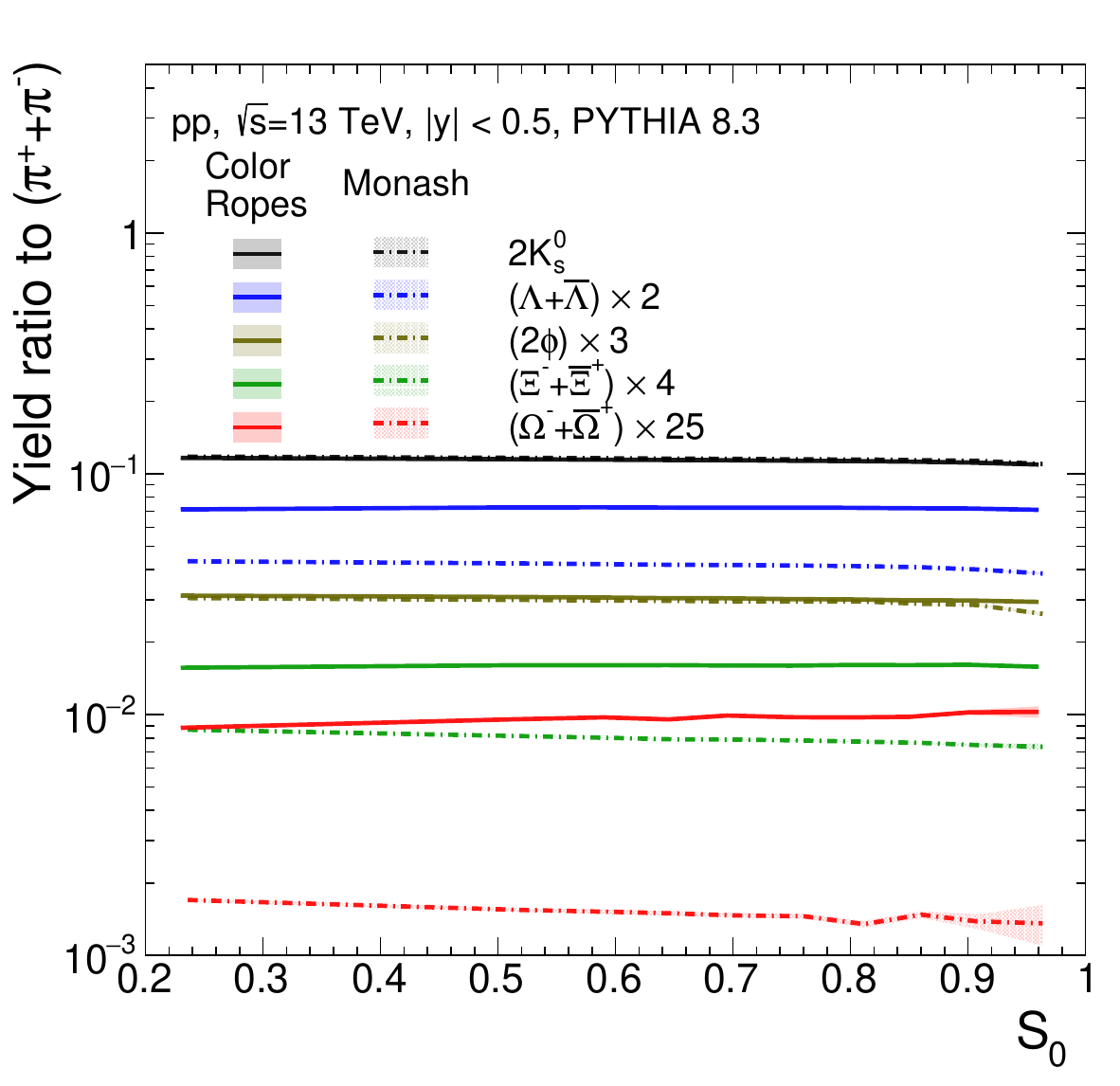}
\includegraphics[scale=0.4]{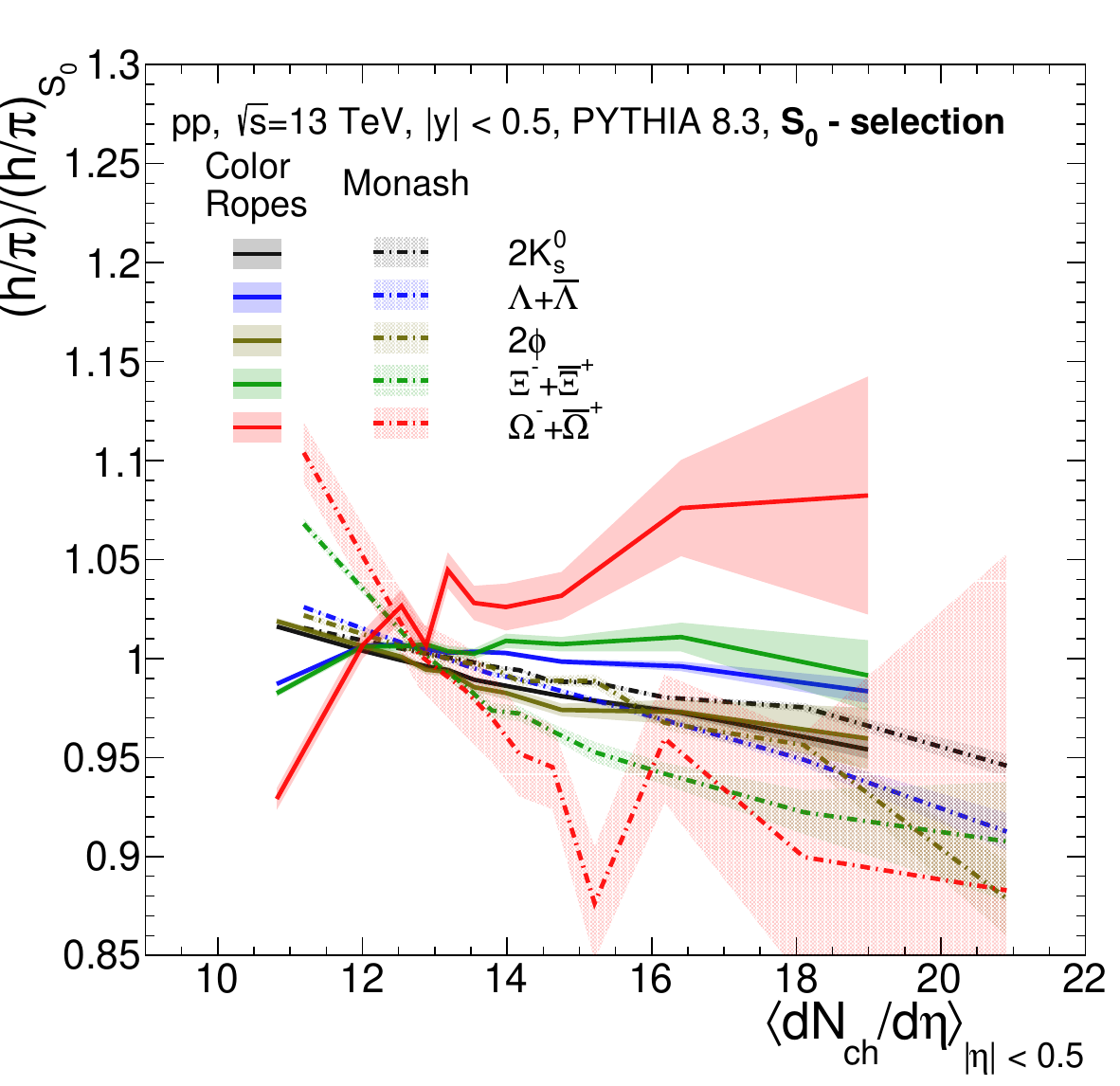}
\caption{$p_{\rm T}$-integrated yield ratios to pions as a function of $p_{\rm T}-$weighted transverse spherocity ($S_{0}$) (top) and particle yield ratios to pions normalized to the values obtained for the spherocity integrated events (bottom) for $|y|<0.5$ in pp collisions at $\sqrt{s}=13$ TeV using Color Ropes and Monash tunes of PYTHIA~8.}
\label{fig:strangespherocity}
\end{center}
\end{figure}

\begin{figure}
\begin{center}
\includegraphics[scale=0.4]{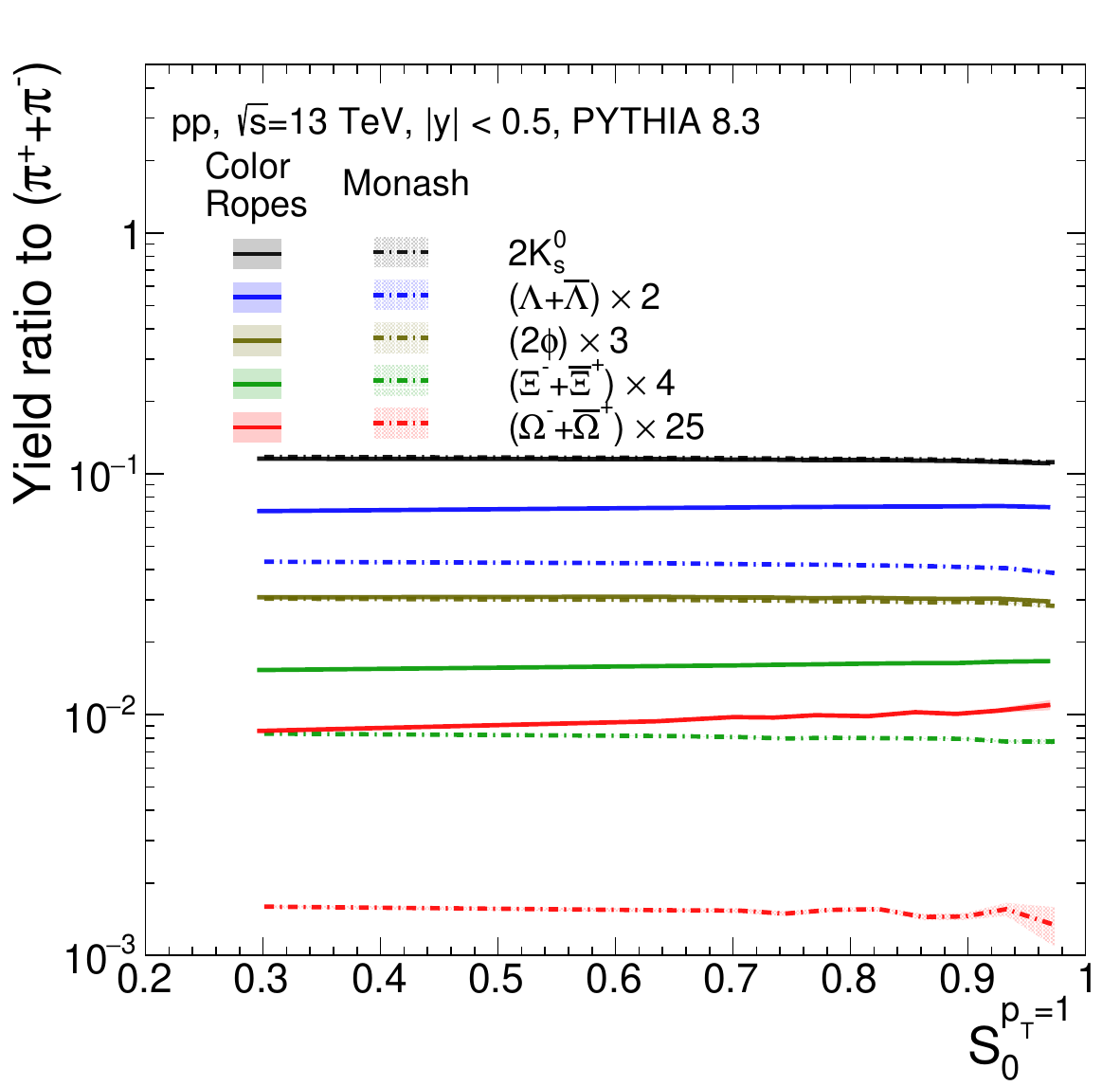}
\includegraphics[scale=0.4]{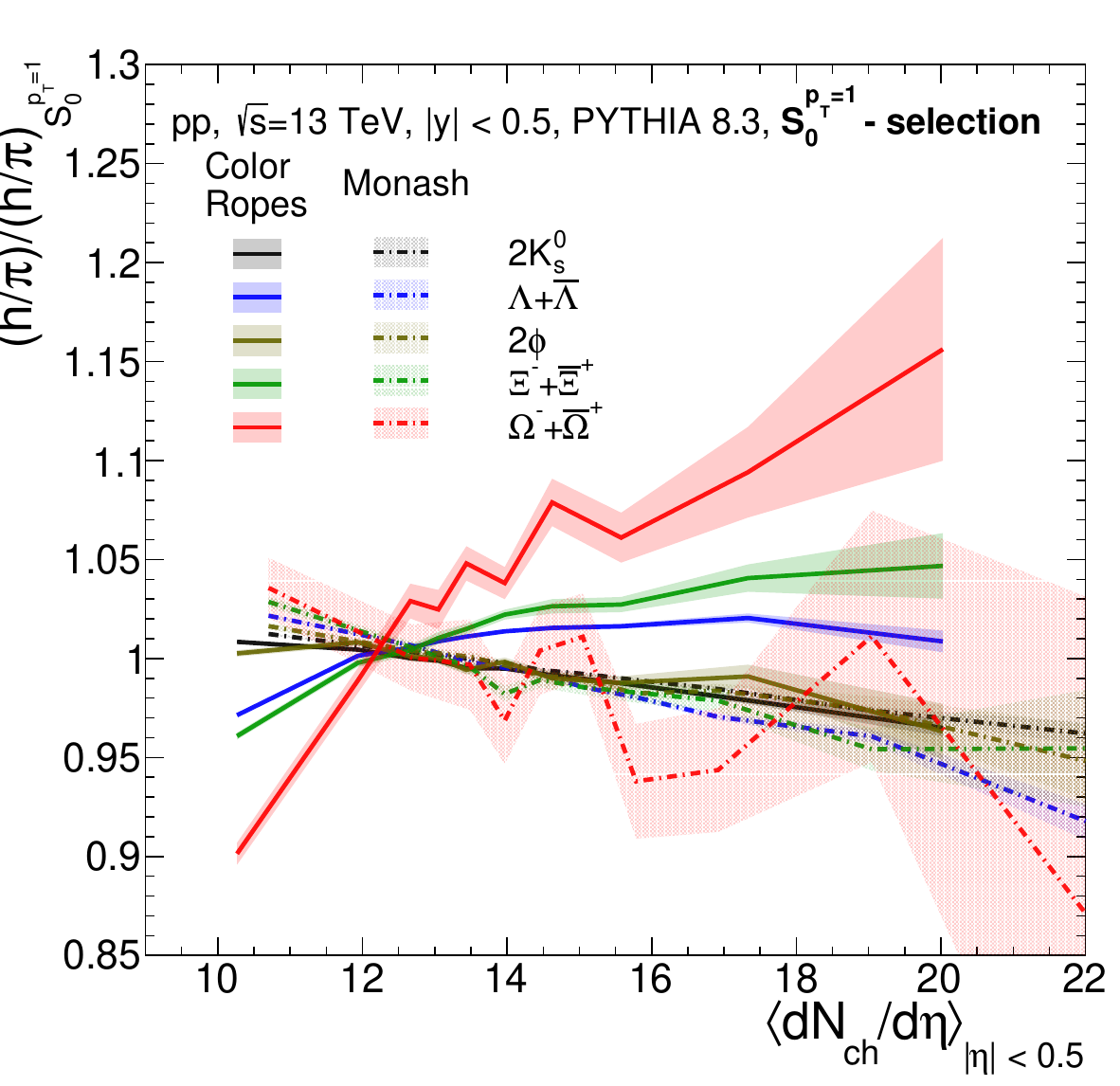}
\caption{$p_{\rm T}$-integrated yield ratios to pions as a function of $p_{\rm T}-$unweighted transverse spherocity ($S_{0}$) (top) and particle yield ratios to pions normalized to the values obtained for the spherocity integrated events (bottom) for $|y|<0.5$ in pp collisions at $\sqrt{s}=13$ TeV using Color Ropes and Monash tunes of PYTHIA~8.}
\label{fig:SNYieldS0}
\end{center}
\end{figure}

The top panels of Figs.~\ref{fig:strangespherocity} and~\ref{fig:SNYieldS0} show the integrated yield ratios of strange and multi-strange hadrons to pions as a function of  $p_{\rm T}-$weighted transverse spherocity ($S_{0}$) and unweighted transverse spherocity ($S_{0}^{p_{\rm T}=1}$) in pp collisions at $\sqrt{s}=13$ TeV using Color Ropes and Monash tunes of PYTHIA~8, respectively. As discussed in the previous section, experimentally, the original $p_{\rm T}-$weighted $S_{0}$ estimator introduces a neutral jet bias and detector smearing effect, reported in Ref~\cite{ALICE:2023bga}. This issue could be fixed by modifying the definition via a new estimator called $p_{\rm T}$-unweighted transverse spherocity estimator ($S_{0}^{p_{\rm T} = 1}$). It is observed that the $K_{S}^{0}/\pi$, $\Lambda/\pi$, $\Xi/\pi$, and $\phi/\pi$ ratios remains almost constant with $S_{0}$ and $S_{0}^{p_{\rm T}=1}$ for the Color Ropes case. In contrast, $\Omega/\pi$ shows small dependence on both $S_{0}$ and $S_{0}^{p_{\rm T}=1}$, which increases around 15\% from lowest to highest classes of transverse spherocity. However, for PYTHIA~8 Monash case, $\Xi/\pi$ and $\Omega/\pi$ shows a decreasing trend as a function of $S_{0}$ and remains nearly constant with increase in $S_{0}^{p_{\rm T}=1}$. It is important to note that, here, the yield ratios of strange hadrons to pions are comparable to the values in high multiplicity events as observed in Figs.~\ref{fig:strangempi}, and~\ref{fig:strangeNch}. This is the consequence of event selection cut of $N_{\rm ch}^{\rm mid}\geq10$ which is applied during the definitions of $S_{0}$ and $S_{0}^{p_{\rm T}=1}$.

The bottom panel of Figs.~\ref{fig:strangespherocity} and~\ref{fig:SNYieldS0} show self-normalized yield ratios of the strange and multi-strange particle to pions as a function of $\langle dN_{\rm ch}/d\eta\rangle$ obtained in the different classes of $S_{0}$ and $S_{0}^{p_{\rm T}=1}$ in pp collisions at $\sqrt{s}=13$ TeV using Color Ropes and Monash tunes of PYTHIA~8, respectively.
In Color Ropes, the self-normalised yield ratio of $\Omega$ to pions has a steeper increase with $\langle dN_{\rm ch}/d\eta\rangle$, when the events are selected with $S_{0}^{p_{\rm T}=1}$ as compared to $S_{0}$. For $\Xi$ and $\Lambda$, $S_{0}$ provides similar double ratio values while a distinction is observed for the $S_{0}^{p_{\rm T}=1}$ case. Here, for $S_{0}^{p_{\rm T}=1}$, the ratio for $\Xi$ increases rapidly in comparison to $\Lambda$ with the increase in $\langle dN_{\rm ch}/d\eta\rangle$, holding onto the scaling of the strangeness quantum number of strange baryons. Furthermore, for $\phi$ and $K^{0}_{S}$, one finds that the double ratios decrease with increase in $\langle dN_{\rm ch}/d\eta\rangle$ for both $S_{0}$ and $S_{0}^{p_{\rm T}=1}$ cases. However, the rate of decrease of the slope is larger for $S_{0}$ case as compared to $S_{0}^{p_{\rm T}=1}$. These trends for $\phi$ and $K^{0}_{S}$ are believed to be a consequence of measuring spherocity and the particles in overlapping rapidity regions, which is absent when event shapes are obtained in different rapidity regions, for example in Figs.~\ref{fig:SNYieldNch} and ~\ref{fig:strangempi}.
In contrast, for PYTHIA~8 Monash case, the self-normalised yield ratios of all the strange hadrons to pions decrease with increase in charged particle multiplicity at midrapidity for both $S_{0}$ and $S_{0}^{p_{\rm T}=1}$ cases similar to the results shown in Fig.~\ref{fig:strangeNch}, where the events are selected with $N_{\rm ch}^{\rm mid}$. This clearly indicates that, for the PYTHIA~8 Monash case, the self-normalised yield ratios of strange hadrons to pions have a dominating effect from the autocorrelation bias when the production yield ratios of strange hadrons to pions are weakly dependent upon $N_{\rm mpi}$ in the absence of RH. Further, the fall of self-normalised yield ratios to pions is weaker when events are selected with $S_0^{p_{\rm T}=1}$ as compared to $S_0$. This confirms that, for the study of strangeness production in pp collisions, $S_{0}^{p_{\rm T}=1}$ possess a small selection bias but a better observable as compared to $S_{0}$ with wider multiplicity coverage.


\begin{figure}
\begin{center}
\includegraphics[scale=0.4]{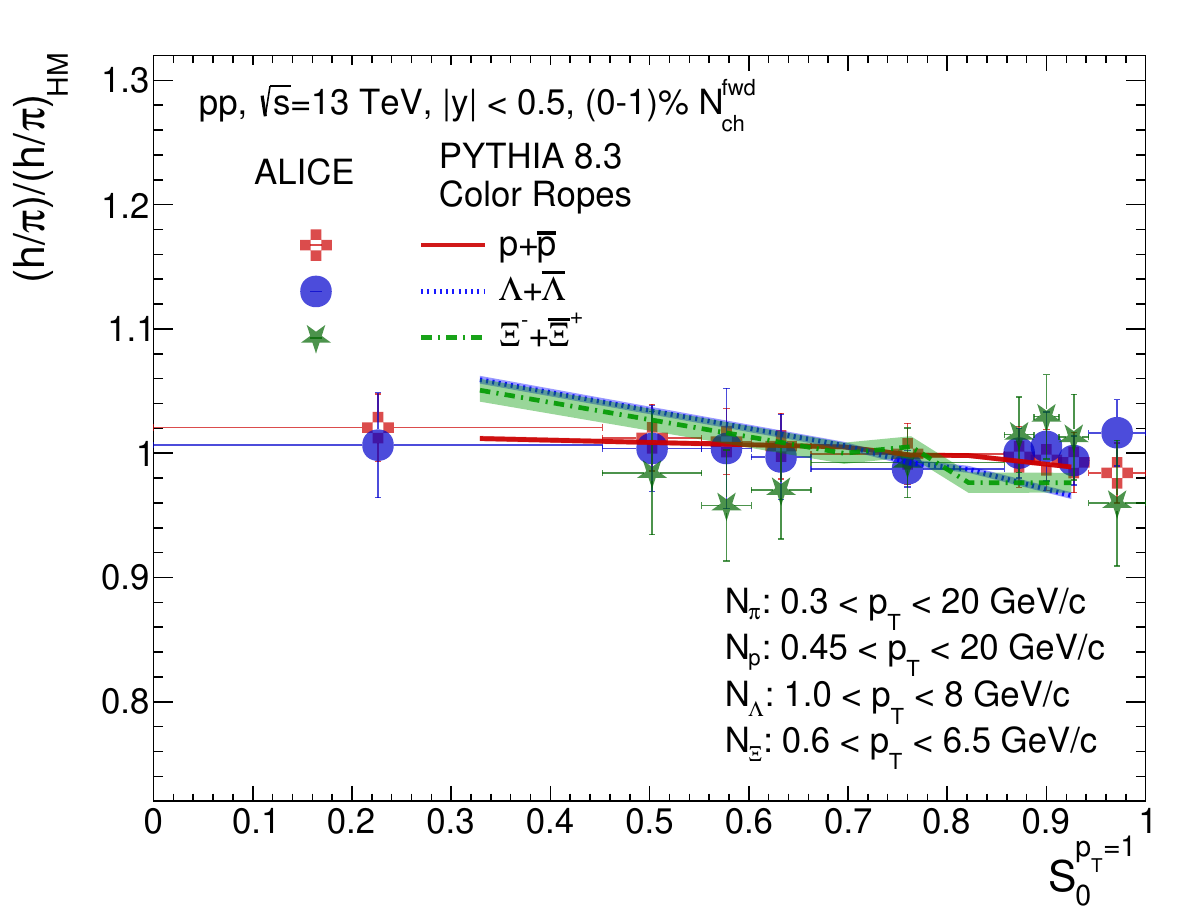}
\caption{Particle yield ratios to pions normalized to the values obtained for the spherocity integrated events for top 1\% $N_{\rm ch}^{\rm fwd}$ events as a function of $S_{0}^{p_{\rm T}=1}$ in pp collisions at $\sqrt{s}=13$ TeV using PYTHIA~8 Color Ropes. The model predictions are compared with the corresponding experimental data with ALICE~\cite{ALICE:2023bga}.}
\label{fig:SNYieldS0HM}
\end{center}
\end{figure}

Figure~\ref{fig:SNYieldS0HM} shows the $p_{\rm T}$-integrated yield ratios to pions normalized to spherocity-integrated events for top 1\% events as a function of $S_{0}^{p_{\rm T}=1}$ in pp collisions at $\sqrt{s}=13$ TeV using PYTHIA~8 Color Ropes and is compared with the corresponding experimental measurements at ALICE~\cite{ALICE:2023bga}. It is observed that both the model predictions and experimental measurements agree with each other and do not show significant dependence on the strangeness content of the particle.

\begin{figure}
\begin{center}
\includegraphics[scale=0.4]{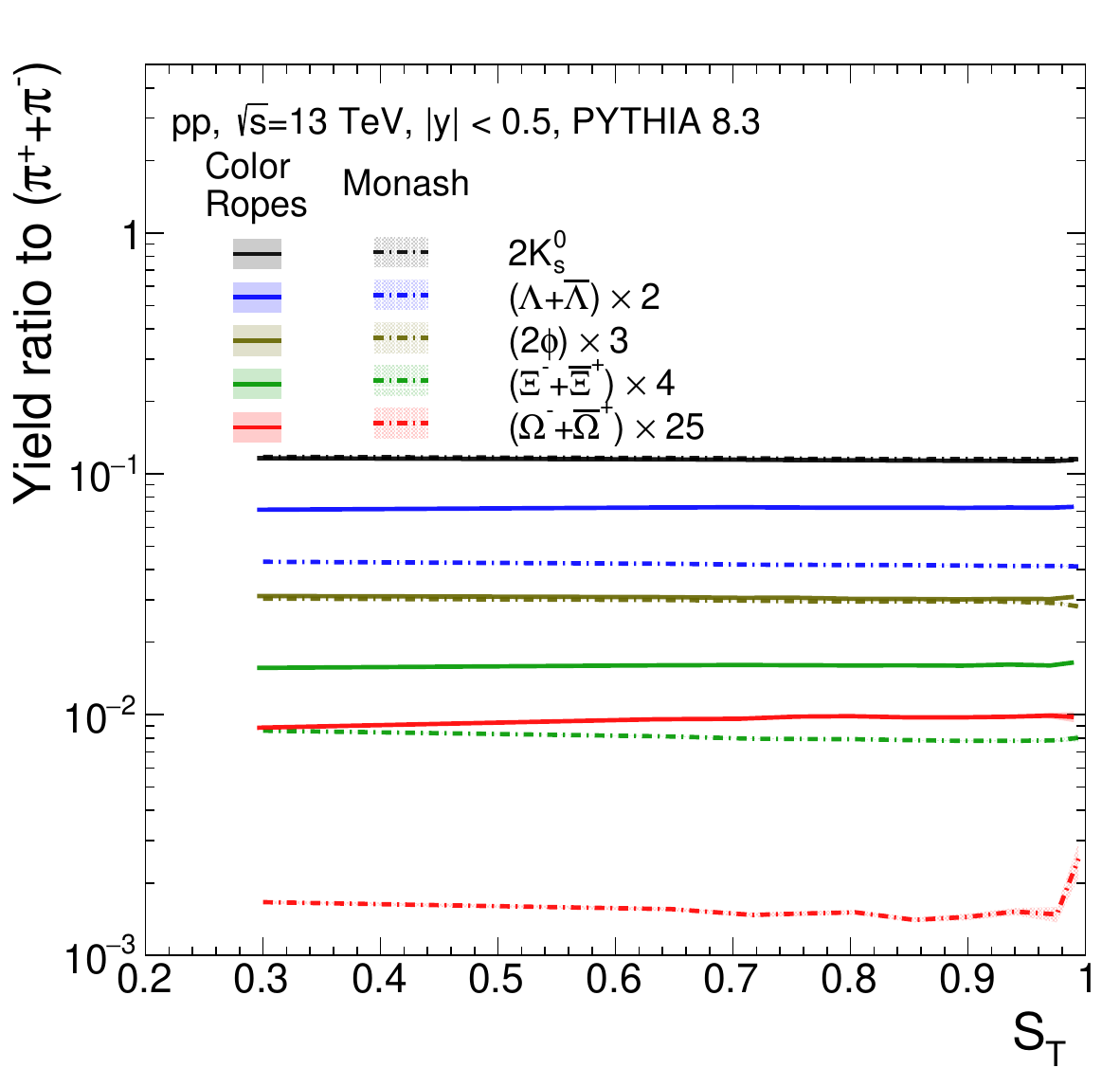}
\includegraphics[scale=0.4]{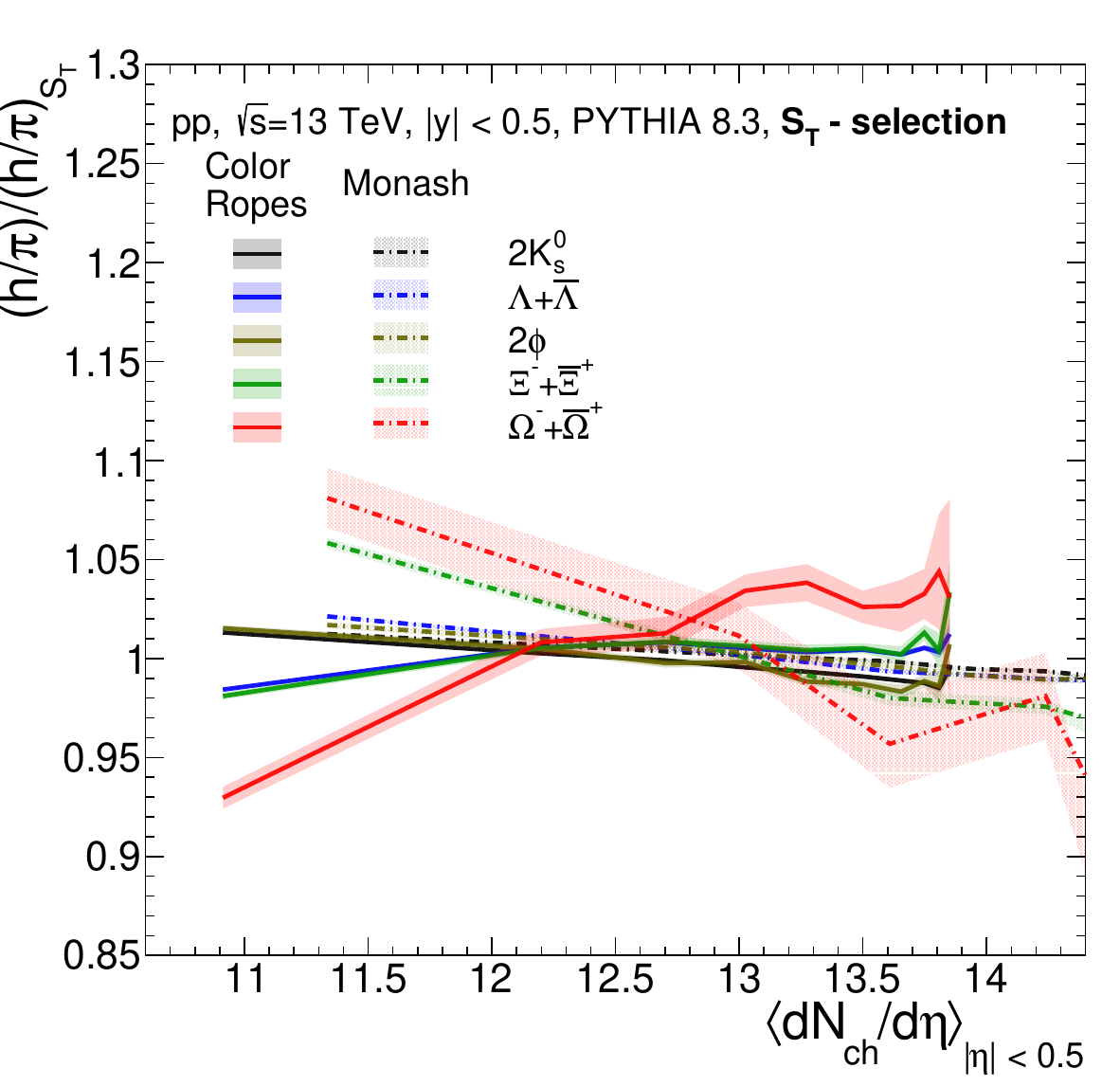}
\caption{$p_{\rm T}$-integrated yield ratios to pions as a function of $p_{\rm T}-$unweighted transverse sphericity ($S_{T}$) (top) and particle yield ratios to pions normalized to the values obtained for the sphericity integrated events (bottom)  obtained for $|y|<0.5$ in pp collisions at $\sqrt{s}=13$ TeV using PYTHIA~8 Color Ropes and Monash.}
\label{fig:strangesphericity}
\end{center}
\end{figure}

\begin{figure*}
\begin{center}
\includegraphics[scale=0.4]{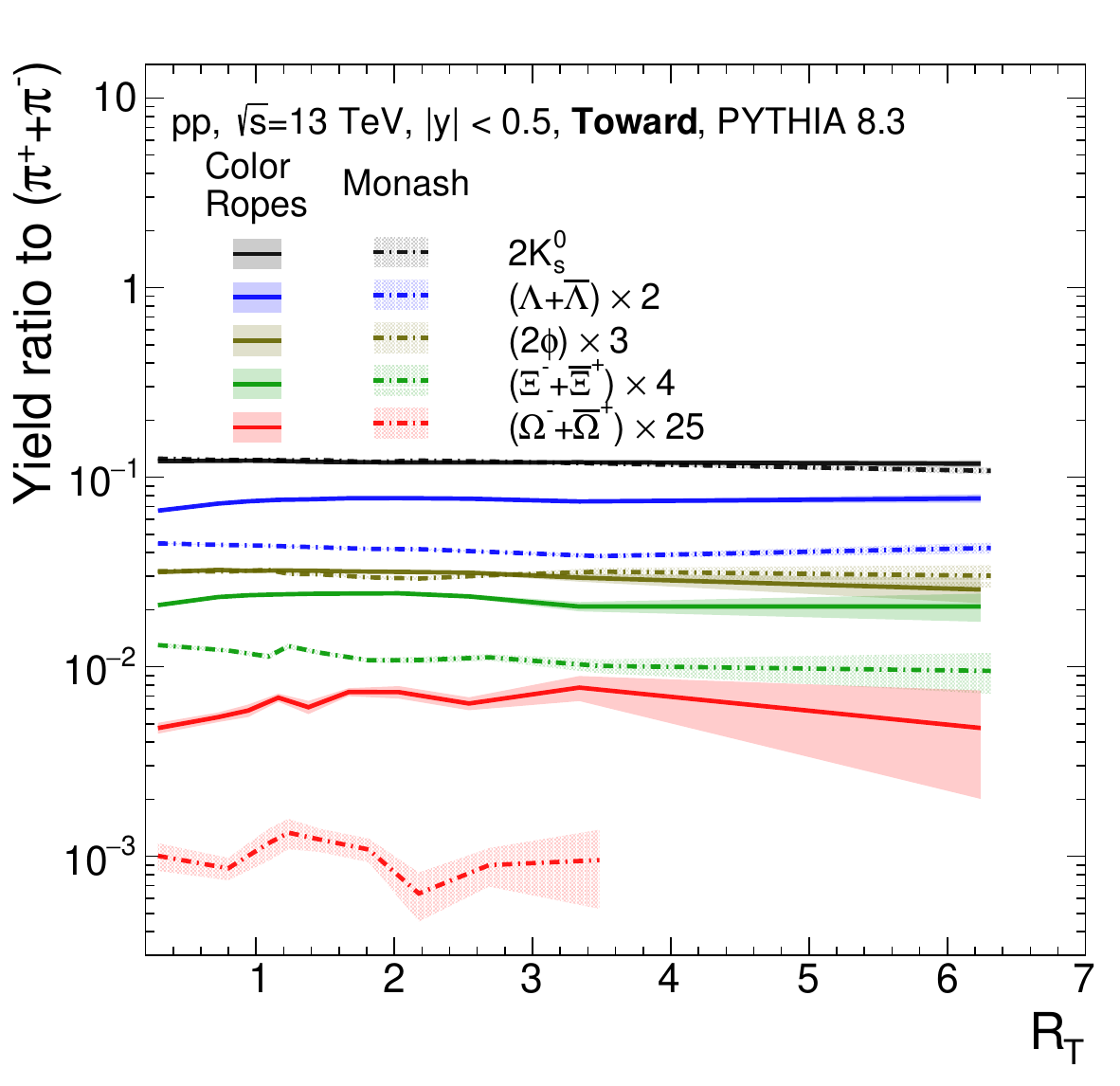}
\includegraphics[scale=0.4]{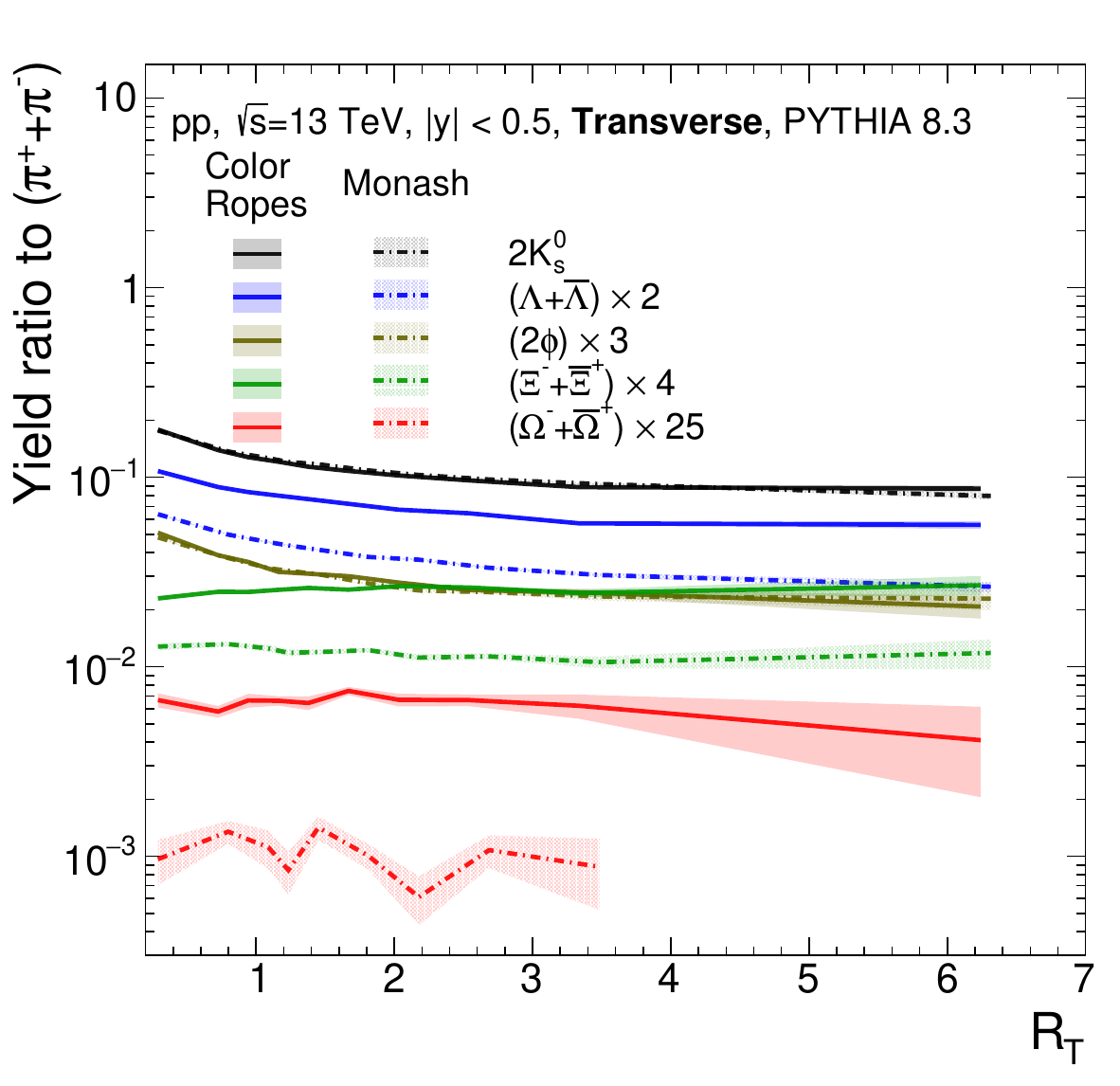}
\includegraphics[scale=0.4]{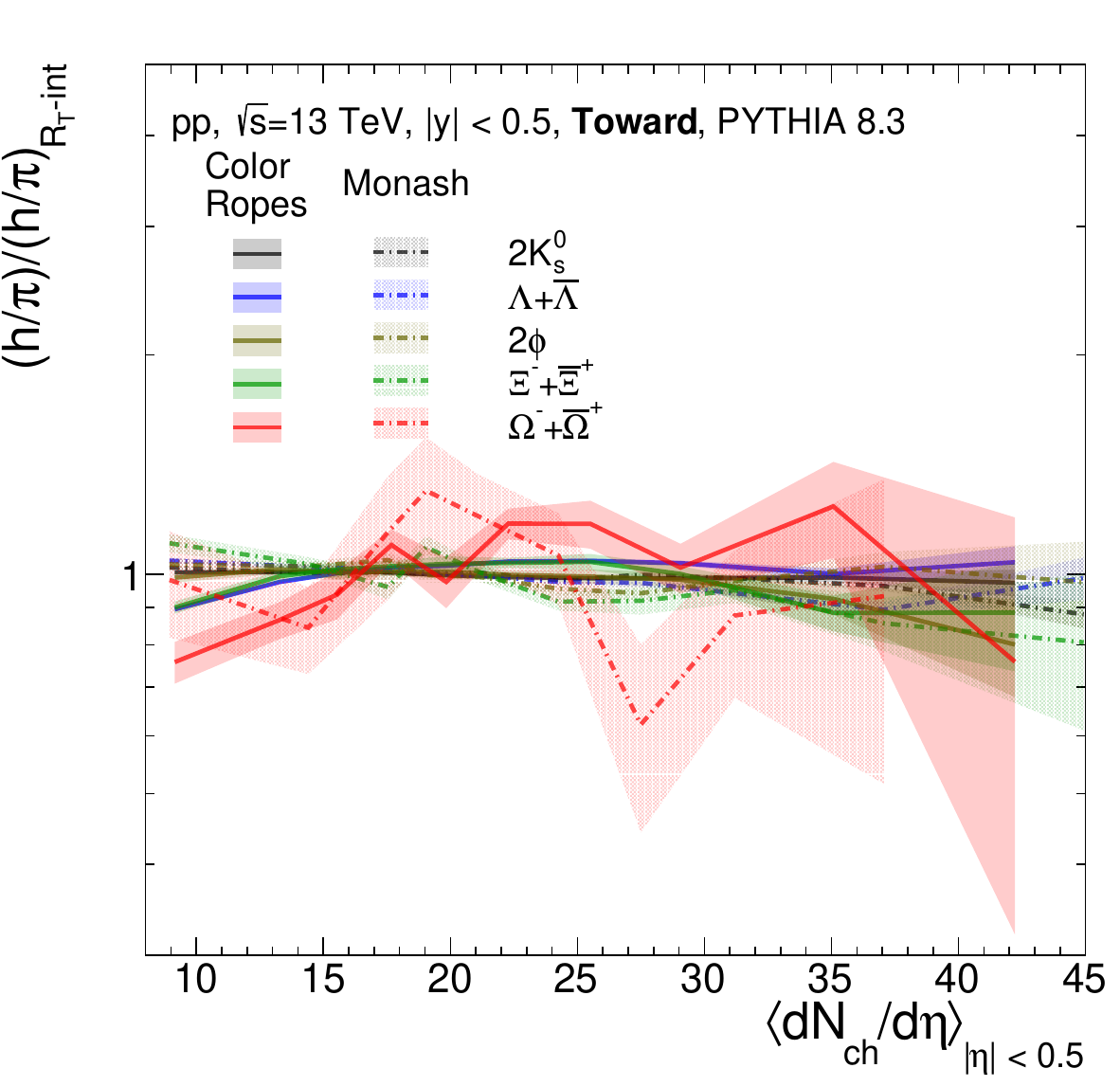}
\includegraphics[scale=0.4]{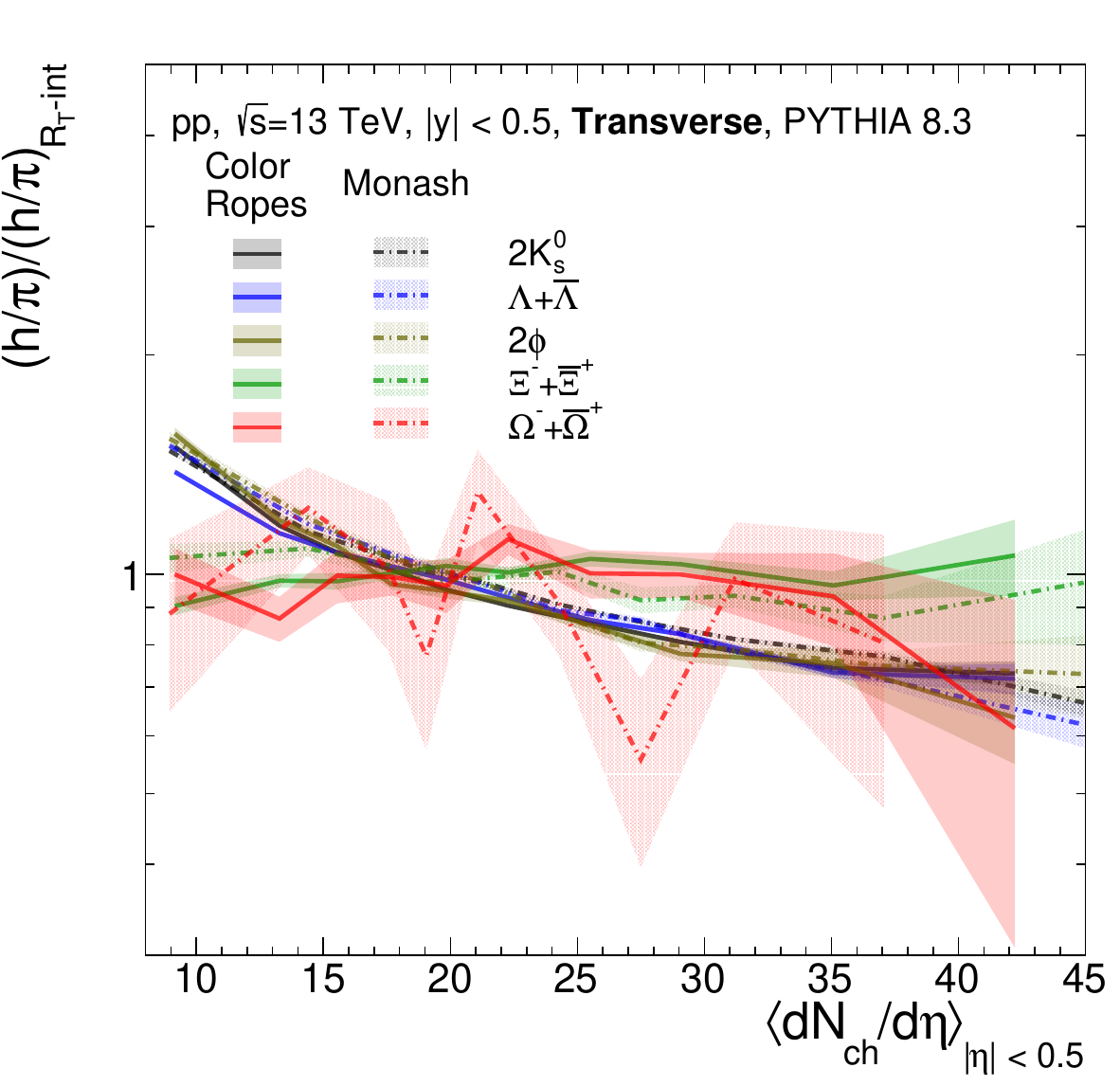}
\caption{$p_{\rm T}$-integrated yield ratio to pions (top) and particle yield ratios to pions normalized to the values obtained for the minimum bias event (bottom) obtained in $|y|<0.5$ as a function of $R_{\rm T}$ in toward (left) and transverse (right) regions of pp collisions at $\sqrt{s}=13$ TeV using PYTHIA~8 Color Ropes and Monash.}
\label{fig:strangeRT}
\end{center}
\end{figure*}

The top panel of Fig.~\ref{fig:strangesphericity} shows the $p_{\rm T}$-integrated yield ratios of strange and multi-strange hadrons to pions as a function of transverse sphericity in pp collisions at $\sqrt{s}=13$ TeV using PYTHIA~8 Color Ropes and Monash. Similar to Fig.~\ref{fig:strangespherocity}, one finds negligible dependence of $K_{S}^{0}/\pi$, $\Lambda/\pi$, $\Xi/\pi$, $\phi/\pi$, and $\Omega/\pi$ on transverse sphericity ($S_{\rm T}$) for PYTHIA~8 Color Ropes. However, PYTHIA~8 Monash results show a decreasing trend for $\Xi/\pi$, and $\Omega/\pi$ as a function of $S_{\rm T}$, similar to observations made in Figs.~\ref{fig:strangeNch} and \ref{fig:strangespherocity}. So, in conclusion, we find that events with isotropic emission of particles have almost similar behavior of strange to non-strange ratios compared to events dominated by jets when we choose events having $N_{\rm ch}^{\rm mid}\geq 10$. Therefore, the transverse sphericity and the weighted and unweighted transverse spherocity event classifiers are inadequate to study the strangeness suppression characteristics in low multiplicity pp collisions using the PYTHIA~8 model; however, their applicability in the highest multiplicity classes is worth investigating.

The bottom panel of Fig.~\ref{fig:strangesphericity} shows strange and multi-strange hadron yield ratios to pions scaled to the values obtained in the sphericity integrated events as a function of $\langle dN_{\rm ch}/d\eta\rangle$ obtained in different classes of $S_{\rm T}$ in pp collisions at $\sqrt{s}=13$ TeV using PYTHIA~8 Color Ropes and Monash. 
Here, similar to $S_{0}$ in Fig.~\ref{fig:SNYieldS0}, one finds similar enhancement trends for $\Omega$, $\Xi$, and $\Lambda$ baryons and suppression trends for $\phi$, and $K^{0}_{S}$ mesons with an increase in $\langle dN_{\rm ch}/d\eta\rangle$ obtained in different classes of $S_{\rm T}$ for PYTHIA~8 Color Ropes. For the PYTHIA~8 Monash case, the self-normalised yield ratios of all the strange hadrons to pions decrease with an increase in charged particle multiplicity density at midrapidity and are stronger for hadrons with having a larger number of valence strange quarks. Notably, the amplitudes for these enhancement or suppression trends using $S_{\rm T}$ are the smallest compared to other event shapes. 


In the top panels of Fig.~\ref{fig:strangeRT} show the $K_{S}^{0}/\pi$, $\Lambda/\pi$, $\Xi/\pi$, $\phi/\pi$, and $\Omega/\pi$ ratios as a function of $R_{\rm T}$ in toward (left) and transverse (right) regions of pp collisions at $\sqrt{s}=13$ TeV using PYTHIA~8 Color Ropes and Monash. Here, we have restricted the strange hadrons $p_{\rm T}<5$ GeV/c to avoid the bias caused by the trigger particle selection for $R_{\rm T}$~\cite{Bencedi:2021tst, Palni:2020shu}. It is observed that the behavior of strange particle ratios is different in toward and transverse regions. In the toward region, where auto-correlation bias is absent, the $\Omega/\pi$ shows a mild increasing trend with respect to increase in $R_{\rm T}$ with PYTHIA~8 Color Ropes. The $\Xi/\pi$ and $\Lambda/\pi$, unexpectedly does not show significant dependence on $R_{\rm T}$. 
However, in the transverse region, where autocorrelation bias is dominant as $R_{\rm T}$ is defined in the same region, the mild $\Omega/\pi$ enhancement disappears in the ratios of strange mesons to pions. In contrast, the neutral hadrons to pion ratios, such as $\phi/\pi$, $K^{0}_{S}/\pi$ and $\Lambda/\pi$ decrease with an increase in $R_{\rm T}$. This effect seems to be coming purely from autocorrelation bias. Similar to sphericity and spherocity, the strangeness enhancement feature is also found to be weak in the relative transverse activity classifier. This could be because of the high-$p_{\rm T}$ selection of trigger particle, one always looks at high multiplicity events, and the multiplicity coverage is similar to spherocity selection, as can be seen in the bottom panels of Fig.~\ref{fig:strangeRT}. The bottom panel shows the strange and multi-strange hadron yield ratios to pions normalized to that for $R_{\rm T}$-integrated events as a function of $\langle dN_{\rm ch}/d\eta\rangle$ in different classes of events selected via $R_{\rm T}$ distributions in pp collisions at $\sqrt{s}=13$ TeV using PYTHIA~8. The double ratio of $\phi$, $K^{0}_{S}$ and $\Lambda$ decrease with increase in $\langle dN_{\rm ch}/d\eta\rangle$ when the events are selected for $R_{\rm T}$ indicating the presence of strong autocorrelation bias. 
 
\begin{figure}
\begin{center}
\includegraphics[scale=0.4]{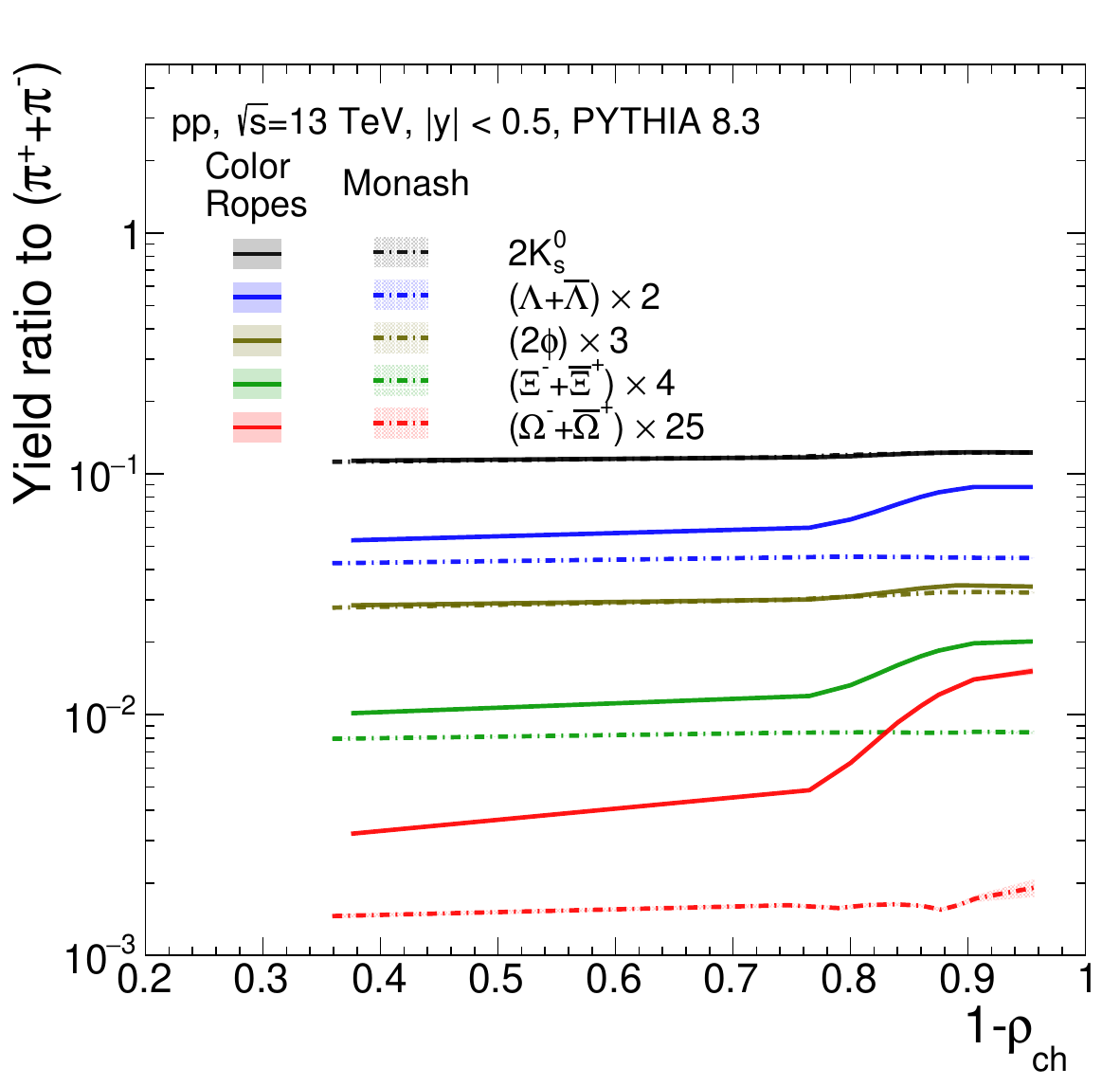}
\includegraphics[scale=0.4]{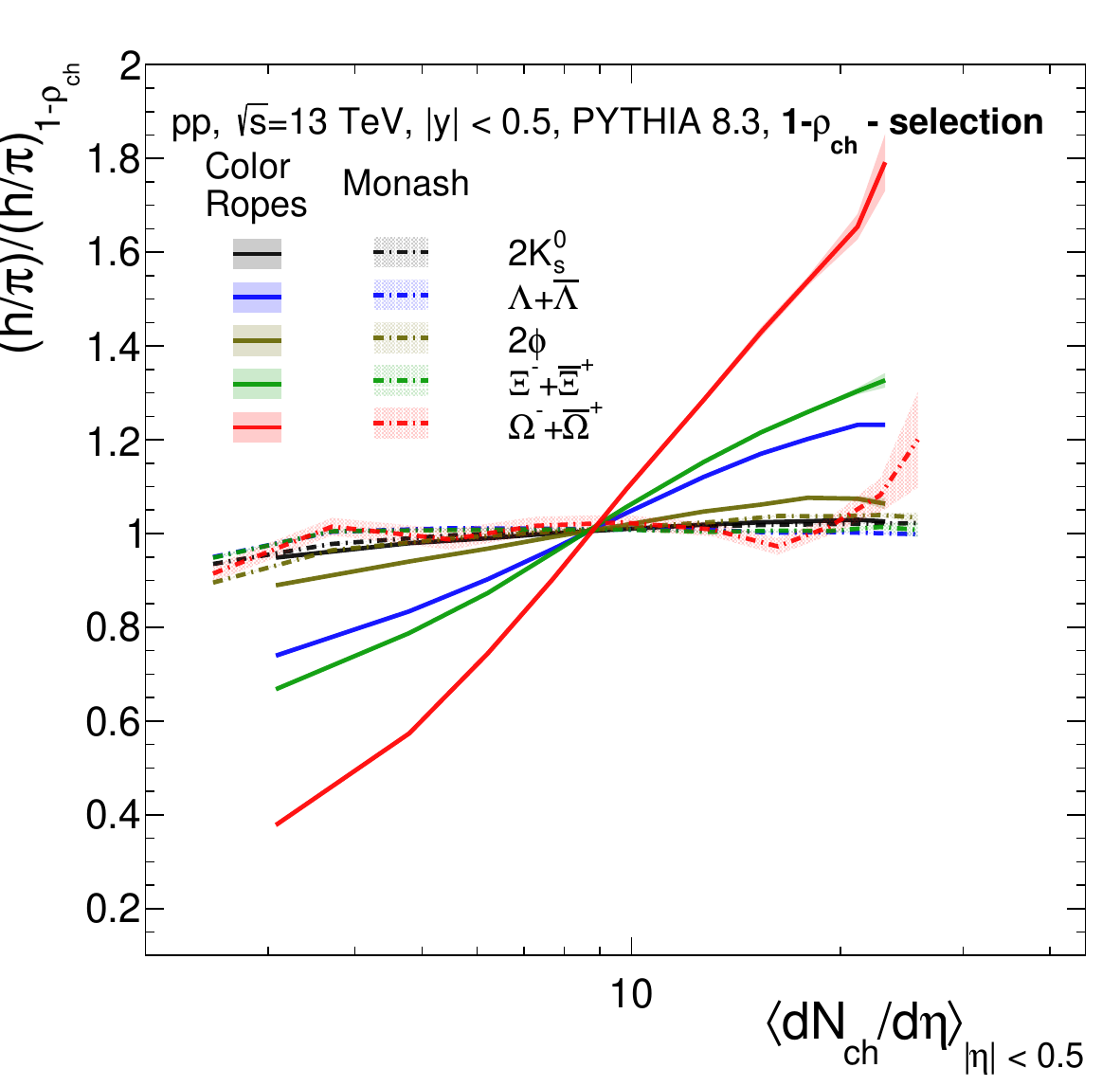}
\caption{$p_{\rm T}$-integrated yield ratio to pions (top) and particle yield ratios to pions normalized to the values obtained for the minimum bias events (bottom) in $|y|<0.5$ as a function of charged particle flattenicity ($\rho_{\rm ch}$) in pp collisions at $\sqrt{s}=13$ TeV using PYTHIA~8 Color Ropes and Monash.}
\label{fig:strangeflat}
\end{center}
\end{figure}

The charged particle flattenicity dependence of the ratios $K_{S}^{0}/\pi$, $\Lambda/\pi$, $\Xi/\pi$, $\phi/\pi$, and $\Omega/\pi$ is shown in top panel of Fig.~\ref{fig:strangeflat} for pp collisions at $\sqrt{s}=13$ TeV using PYTHIA~8 Color Ropes and Monash. It is interesting to note that these ratios increase as a function of 1-$\rho_{\rm ch}$ for all considered strange and multi-strange hadrons, except for $K_S^{0}$ for PYTHIA~8 Color Ropes. A sudden increase in the slope of the ratios is observed in Fig.~\ref{fig:strangeflat} for the higher values of $1-\rho_{\rm ch}$ ($0.75 \lesssim 1-\rho_{\rm ch} \lesssim 0.95 $) compared to the lower values of $1-\rho_{\rm ch}$ ($0.35 \lesssim 1-\rho_{\rm ch} \lesssim 0.75 $). It is found that the rate of increase of the strange to non-strange particle ratios is strange quantum number dependent; the triple-strange baryons such as $\Omega$ have more slope compared to the single and double strange baryons such as  $\Xi$, and $\Lambda$. In addition, the hidden strange meson, $\phi$ shows a small dependence on event selection based on $1-\rho_{\rm ch}$. Here, the values of yield ratios of the multi-strange hadrons to pions for the lowest and highest $1-\rho_{\rm ch}$ classes are comparable to corresponding classes of $N_{\rm mpi}$ and $N_{\rm ch}^{\rm fwd}$ in Figs.~\ref{fig:SNYieldNch} and~\ref{fig:strangempi}, respectively. We do not observe any change in the yield ratios of strange hadrons to pions as a function of $1-\rho_{\rm ch}$.

The bottom panel of Fig.~\ref{fig:strangeflat} shows the strange hadron yield ratios to pions scaled to those obtained in flattenicity integrated events as a function of $\langle dN_{\rm ch}/d\eta\rangle$ obtained in different classes of (1-$\rho_{\rm ch}$) in pp collisions at $\sqrt{s}=13$ TeV using PYTHIA~8 Color Ropes and Monash. 
For PYTHIA~8 Color Ropes, the double ratios for all strange and multi-strange hadrons increase linearly with $\log(\langle dN_{\rm ch}/d\eta\rangle)$. The rate of increase of the double ratio with $\log(\langle dN_{\rm ch}/d\eta\rangle)$ is highest for $\Omega$, while the slope decreases and becomes almost zero for $K^{0}_{S}$. The hidden strange meson, $\phi$, lies between $K^{0}_{S}$ and $\Lambda$, which is close to the observations in Fig.~\ref{fig:strangempi} where events are selected with $N_{\rm mpi}$. The close resemblance between strange hadron production with $N_{\rm mpi}$ and $\rho_{ch}$ makes charged particle flattenicity one of the ideal choices among the existing event shapes for the study of strangeness production in pp collisions at the LHC.


\section{Summary}
\label{summary}
We present an extensive summary of the strange particle ratios to pions as a function of different event classifiers using the PYTHIA~8 model with color reconnection and rope hadronization mechanisms to understand the microscopic origin of strangeness enhancement in pp collisions and also prescribe the applicability of these event classifiers in the context of strangeness enhancement. Comparisons with default Monash with and without color reconnections have also been made. The event shape observables used in this study are chosen based on their different coverages in terms of $\langle N_{\rm mpi} \rangle$, $\langle \hat{p}_{\rm T}\rangle$ and $\langle dN_{\rm ch}/d\eta \rangle_{|\eta|<0.5}$. We now summarize the conclusions from each event classifier.

\begin{itemize}
\item{{\bf Number of multi-parton interactions}: MPI plays a pivotal role in particle production, and it is also observed for strange and multi-strange hadrons. A clear strangeness enhancement (suppression) is seen for high (low) multiplicity pp collisions when studied as a function of the number of multi-partonic interactions for PYTHAI~8 color ropes. This suggests that RH is necessary to show the strangeness enhancement feature in PYTHIA~8, where the role of MPI is to increase the number of color ropes formed.}

\item{\textbf{Charged particle multiplicity in mid-rapidity}: Multiplicity measured in mid-rapidity shows strangeness enhancement, but the results are significantly prone to autocorrelation bias. The strangeness suppression in low multiplicity is not seen except for $\Omega$ in Color Ropes. }

\item{\textbf{Charged particle multiplicity in forward-rapidity}: Multiplicity measured in forward-rapidity significant reduction of autocorrelation bias, but it is prone to a selection bias of choosing high-momentum particles. PYTHIA~8 Color Ropes describes the observed strangeness enhancement by the ALICE experiment.}

\item{\textbf{Transverse spherocity}: Due to an implicit cut on the minimum number of charged particles being larger than 10, the multiplicity reach for the studies based on spherocity selection is limited towards high multiplicity. Even at high multiplicity, the isotropic events show larger enhancement compared to jetty events. The usage of $p_{\rm T}$-unweighted definition of spherocity definition results in larger enhancement in high multiplicity collisions.}

\item{\textbf{Transverse sphericity}: A similar behavior of spherocity is seen when studied as a function of transverse sphericity. However, quantitatively the strangeness enhancement is lesser compared to transverse spherocity.}

\item{\textbf{Relative transverse activity classifier}: With a high transverse momentum cut for trigger particle, one probes high multiplicity events with $R_{\rm T}$. The studies based on $R_{\rm T}$ in the transverse region show significant autocorrelation bias while towards the region, which is free from biases, shows mild strangeness enhancement for $\Omega$.}

\item{\textbf{Charged particle flattenicity:} The most recent event shape observable, charged particle flattenicity, is found to be most suited for the study of strangeness-enhancement and at high multiplicity. It shows a similar quantitative enhancement as seen for the study-based on number of multi-parton interactions. }

\end{itemize}

As, significantly higher statistics would be available in Run 3 of the LHC with respect to Run 1 and Run 2, all the above discussed event-classifiers can be experimentally used to probe strangeness with a high level of precision. This paper provides a baseline for such studies and the comparison of experimental results will give a clear insight into the microscopic origin of strangeness enhancement (suppression) in high (low) multiplicity pp collisions at the LHC.

\section{Acknowledgment}
S.P. acknowledges the doctoral fellowships
from the University Grants Commission (UGC), Government
of India. S.T. acknowledges the CERN Research Fellowship. The authors acknowledge the DAE-DST,
Government of India funding under the Mega-Science
Project—“Indian participation in the ALICE experiment
at CERN” bearing Project No. SR/MF/PS-02/2021-
IITI (E-37123).

\end{document}